# Fundamental Thermo–Visco Mechanical Interactions Governing the Acoustic Response of Laser-Excited Nanoparticles

Stefano Giordano,[1] Michele Diego,[2] Francesco Banfi,[3] and Michele Brun[4]

[1] CNRS, Centrale Lille, Univ. Polytechnique Hauts-de-France, UMR 8520 -IEMN- Institut d'Électronique de Microélectronique et de Nanotechnologie, University of Lille, F-59652 Villeneuve D'Ascq, France

[2] Institute of Industrial Science, The University of Tokyo, Tokyo 153-8505, Japan

[3] Université Claude Bernard Lyon 1, CNRS, Institut Lumière Matière, UMR5306, F-69622, Villeurbanne, France

[4] Department of Mechanical, Chemical and Materials Engineering, Università degli Studi di Cagliari, 09123 Cagliari, Italy

(*Electronic mail: stefano.giordano@univ-lille.fr)

(Dated: 5 March 2026)

In this work, we investigate the thermoacoustic generation and propagation of spherical waves in a viscous fluid induced by a laser-heated spherical particle. Periodic laser excitation gives rise to two coupled mechanisms of acoustic emission. Heat transfer from the particle to the surrounding fluid produces periodic compressions and rarefactions, giving rise to the thermophone effect, while periodic thermal expansion of the solid particle modulates its radius and launches acoustic waves through a piston-like action, known as the mechanophone effect. The thermophone contribution dominates at low frequencies, whereas the mechanophone mechanism becomes more relevant at higher frequencies, with the crossover governed by the interfacial thermal resistance at the solid–fluid boundary. We investigate the effect of nanoparticle embedding fluid viscosity on acoustic wave propagation. Viscous dissipation has a significant impact on attenuation and substantially alters the acoustic penetration depth, thereby affecting the effectiveness of the signal transmission. Viscous damping plays a key role in the mechanophone effect, where hypersonic frequency waves are generated, notably by photoacoustic excitation with picosecond and subpicosecond laser pulses. We develop a theoretical model based on the coupled conservation equations of mass, momentum, and energy in both phases, explicitly accounting for thermal diffusion and viscous losses. The reciprocal coupling between thermal and acoustic fields is fully described, allowing us to quantify how frequency and fluid viscosity jointly control the penetration length of the generated acoustic waves in realistic media. Finally, we discuss the implications for theranostics, highlighting how ensembles of laser-activated particles embedded in biological tissue may be optimized for diagnostic and therapeutic applications.

## I. INTRODUCTION

The generation of acoustic waves in fluids—whether liquid or gaseous—is a fundamental phenomenon underlying a broad range of technological and scientific applications. From traditional sound emission systems to modern, non-invasive imaging and diagnostic tools, acoustic generation plays a central role across disciplines such as biomedicine, industrial non-destructive evaluation, and underwater communication.[1]

Among the various mechanisms for producing acoustic waves, the piezoelectric effect, discovered in 1880,[2] has long been the cornerstone of acoustic transduction technologies. Its foundational principles, established through the works of Lippmann, Voigt, Langevin, and others,[3–8] have supported a century of innovation in both sensing and actuation. Despite the maturity of piezoelectric technology, limitations in frequency bandwidth have motivated the exploration of alternative approaches. In aqueous media, for example, the operational range of piezoelectric transducers typically remains confined to 50–200 MHz, with only a few exceptional realizations extending to about 300 MHz.[9–11]

To overcome these constraints, novel strategies have been introduced, notably capacitive micromachined ultrasonic transducers (CMUTs), where the acoustic signal is produced by the electrostatic attraction between capacitor plates.[12] The issue of bandwidth became particularly pressing with the development of nanoresonators, whose dimensions—on the order of tens of nanometers—naturally give rise to vibrational modes in the hypersonic frequency range.[13,14]

The pursuit of broadband acoustic sources suitable for mesoscopic and nanoscopic scales has consequently revived interest in thermoacoustic mechanisms, in particular the thermophone effect. In its simplest description, a transducer periodically heated while in contact with a fluid induces a temperature field oscillation within the latter. The resulting thermal expansion and contraction generate pressure fluctuations that propagate as sound waves. Because it does not rely on any resonance phenomenon, this mechanism inherently supports a wide frequency spectrum.

Heat can be supplied to the transducer either by electrical driving—where the conversion process is traditionally called the thermophone effect—or by optical illumination, which gives rise to the photophone, photothermal, or photoacoustic effects. Interestingly, both concepts were discovered contemporaneously with the piezoelectric effect but developed independently, forming distinct research branches in acoustic generation: thermophone,[15–19] and photophone.[20–27]

For many decades, however, thermoacoustic generation remained limited in efficiency due to the absence of suitable materials combining high thermal conductivity with low volumetric heat capacity. The advent of nanotechnology has dramatically changed this landscape. New classes of nanostructured materials—exhibiting excellent thermomechanical properties—have enabled efficient thermophone-based transducers.[28–31] Examples include carbon nanotube assemblies,[32–38] metallic micro- and nanowires made of aluminum, gold, or silver,[39–41] as well as graphene films and boron nitride foams.[42–48]

The resurgence of thermoacoustic approaches, often through photoacoustic realizations, has in turn stimulated significant theoretical and modeling efforts aimed at optimizing energy conversion efficiency. Early quantitative models—developed for photoacoustic spectroscopy, microscopy, and imaging—provided a foundational understanding of the coupling between thermal and acoustic fields.[49–52] Notably, the concept of the piston model established a key framework for interpreting the emission dynamics of photothermoacoustic sources.[52]

Subsequent investigations have focused on improving the high-frequency response of the thermophone. It is now well established that the thermal diffusion length in the fluid scales as $\omega^{-1/2}$, where $\omega$ denotes the modulation angular frequency. As a result, at increasing frequencies, the thermal wave becomes confined near the solid–fluid interface, thereby reducing the amplitude of the generated acoustic wave and diminishing the overall efficiency of the process.[53–65]

Recently, a complementary mechanism—termed the mechanophone—has been proposed to explain thermoacoustic generation in this high-frequency regime.[66–68] In contrast to the thermophone, where heat flux into the fluid directly drives compressional waves, the mechanophone effect arises from the thermoelastic oscillations within the solid itself. These internal stress–strain dynamics are mechanically transmitted across the interface, launching acoustic waves into the surrounding medium. Numerical and theoretical studies have made it possible to understand certain aspects of this process in the time domain or for one-dimensional structures.[66–68] However, a complete analytical description is still lacking for spherical nanoparticles, particularly concerning the effects of fluid viscosity. Such a formulation would enable a clear identification of the governing parameters that control the transition between the thermophone and mechanophone regimes, ultimately guiding the rational design of efficient thermoacoustic nanotransducers operating at high frequencies.

This point is crucial since laser-activated particles have emerged as versatile tools for biomedical applications, bridging therapeutic and diagnostic domains through controlled energy conversion at the microscale (see Fig. 1, left panel). When illuminated by an external laser, these particles can efficiently transform optical energy into localized heat or generate acoustic waves via thermoelastic transduction.[69–73] Such dual functionality enables two complementary applications: localized hyperthermia or photothermal therapy for treatment, and photoacoustic or ultrasonic imaging for diagnosis.[74–77]

In therapeutic contexts, localized heating of tissue is used to selectively ablate or weaken pathological cells while minimizing collateral damage to healthy regions. The desired temperature range typically lies between 42–45 °C for mild hyperthermia or exceeds 50 °C for photothermal ablation.[75,78] Gold-based nanoparticles, including nanospheres, nanorods, and nanoshells, are particularly efficient photothermal transducers owing to their strong plasmonic absorption in the near-infrared (NIR) spectral window, where optical penetration in biological tissues is maximized.[79,80] Other materials, such as iron oxide ($Fe_3O_4$), carbon nanotubes, or semiconducting nanostructures, have also been employed for magnetothermal or photothermal treatments.[81] Typical particle diameters range from 5 nm to 700 nm, balancing optical absorption efficiency, tissue penetration, and biocompatibility.[82]

Beyond therapy, these same particles serve as contrast agents. Under pulsed laser excitation, transient heating induces rapid thermoelastic expansion, producing acoustic waves that can be detected to reconstruct high-contrast images of tissue architecture and molecular composition.[83–88] Materials with high optical absorption and low heat capacity—such as gold nanorods, carbon-based nanoparticles, or metal–organic frameworks—enhance the photoacoustic signal intensity by efficiently coupling light absorption to acoustic emission.[89–91] In ultrasound imaging, micro- or nanoparticles can also improve acoustic scattering contrast, providing complementary diagnostic information.[92]

Using localized particles embedded in tissue offers substantial advantages compared to external transducers or heating devices. First, the spatial precision of energy deposition can be greatly enhanced, enabling selective treatment or imaging of specific microregions. Second, internal particles minimize attenuation losses and thermal diffusion, allowing efficient heating or acoustic generation even in optically or acoustically challenging environments. Third, these systems allow multimodal operation, where the same agent can both generate diagnostic photoacoustic signals and provide therapeutic heating—an approach often termed theranostics.[93] Finally, the possibility of targeted delivery—through surface functionalization with antibodies, peptides, or ligands—enables selective accumulation in tumors or diseased tissues, further enhancing treatment specificity.[94]

Nevertheless, several limitations must be considered. Biocompatibility and toxicity remain major concerns, as metallic or inorganic nanoparticles may induce oxidative stress, inflammation, or long-term accumulation in organs such as the liver and spleen.[95,96] The heterogeneous distribution of particles within tissues may lead to uneven heating and reduced therapeutic control. Furthermore, the optical attenuation of tissues restricts laser penetration depth, limiting the effective treatment or imaging volume. Addressing these challenges requires careful design of particle size, coating, and composition to optimize stability, biodistribution, and clearance.[97,98]

Overall, the use of embedded laser-activated particles represents a promising direction for localized and minimally invasive photoacoustic and photothermal therapies. Understanding the underlying thermoacoustic generation mechanisms—including both heat diffusion–driven (thermophone) and particle expansion–driven (mechanophone) effects—is crucial for optimizing their performance and expanding their biomedical applicability.

In this work, we develop a theoretical framework to quantify the thermoacoustic fields generated within and around a single solid laser-excited particle, such as a gold sphere, em-bedded in a surrounding viscous fluid, for example, water (see Fig. 1, right panel). The model simultaneously accounts for the balance equations of mass, linear momentum, and energy in both phases, thereby capturing the full thermomechanical coupling of the system.[99,100] At the solid–fluid interface, we enforce the continuity of the normal mechanical traction, thermal flux, and velocity field. A temperature discontinuity is present, however, due to the effect of an interfacial thermal resistance (also known as Kapitza resistance), which quantifies the impediment to heat transfer across the interface between dissimilar materials (see Fig. 1, right panel).[101–103] This effect can be regarded as one of the possible boundary conditions describing an imperfect interface for transport processes.[104–106] A uniform time-harmonic heating condition at angular frequency $\omega$ is imposed on the particle to represent the energy input from a laser excitation source. The resulting coupled field equations admit an analytical solution, from which the complete temperature, heat flux, velocity, and pressure distributions—both inside the solid particle and in the surrounding fluid—can be explicitly determined. Importantly, the proposed model accounts for the viscosity of the surrounding fluid medium (in addition to that of the spherical particle, whose contribution is comparatively minor). Viscous effects are examined in detail because they play a central role in determining the penetration depth of acoustic waves and, consequently, their potential therapeutic or diagnostic applications. We show how the interplay between frequency and viscosity governs the acoustic penetration length in realistic media.

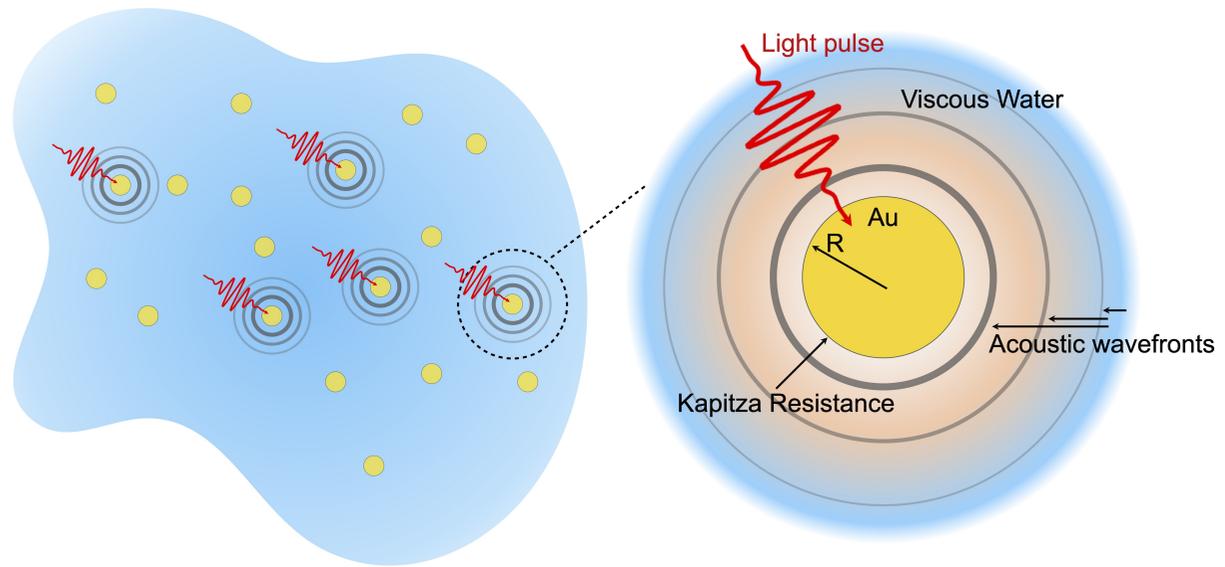

FIG. 1. Left panel: population of spherical gold nanoparticles used for therapeutic and/or diagnostic (theranostic) biomedical applications, embedded in a biological tissue. Right panel: single spherical gold nanoparticle immersed in a viscous fluid (water), activated by a laser light pulse, and generating a thermoacoustic wave in the surrounding environment. The interfacial thermal resistance (also known as Kapitza resistance) regulates the heat flow between particle and fluid by controlling the thermoacoustic generation process.

Frequency-sweep analysis remains a fundamental tool for investigating photoacoustic phenomena induced by pulsed laser excitation. Indeed, the present formulation is developed for monochromatic excitation, namely for a single radiation frequency, and yields the corresponding physical fields throughout the system. This frequency-domain approach provides the fundamental building block for treating more general time-dependent excitations. Indeed, once the incident spectrum is specified—irrespective of its form, including arbitrarily shaped pulses—it can be decomposed into its harmonic components via standard spectral techniques. Each harmonic may then be analyzed independently within our framework, exploiting the linearity of the governing equations. The full response of the system could subsequently be reconstructed by superposing the contributions of all harmonics, weighted according to their spectral amplitudes. Finally, by performing a numerical inverse transform, one can recover the complete time-dependent evolution of the physical fields, thereby bridging the frequency-domain solution with the temporal response of the system. Examples of impulse responses over time can be found in the literature.[67,68] A pulsed laser is composed of harmonic components spanning a frequency range from 0 up to approximately $\tau^{-1}$, where $\tau$ denotes the laser pulse duration. Since the hypersonic acoustic modes associated with the mechanophonic effect extend up to the frequency range of $10^{11}$ rad/sec, the generation of such waves in water requires laser pulses with a duration of the order of picoseconds or less, rather than pulses of the order of nanoseconds. At these ultra-high frequencies, dissipative effects arising from fluid viscosity become crucial and must be explicitly accounted for in the acoustic response.

As mentioned earlier, two distinct phenomena can be identified: at low frequencies, the thermoacoustic response is dominated by the thermophone mechanism, whereas at high frequencies, the mechanophone contribution becomes predominant.[66–68] To demonstrate this behavior, we perform two separate analyses. In the first one, the thermal expansion coefficient of the solid is set to zero so that acoustic waves in the fluid arise solely from thermal compression and expansion within the fluid itself (pure thermophone regime). In the second analysis, the thermal expansion coefficient of the fluid is set to zero, allowing the generation of acoustic waves only through the periodic modulation of the particle size induced by internal thermal stresses (pure mechanophone regime). When both thermal expansion coefficients are assigned their realistic values, the full solution of the coupled problem is recovered. This analysis is carried out over a broad frequency range, enabling us to track how the crossover between the two mechanisms shifts with the variation of the Kapitza resistance. The proposed solution allows for the examination of all thermoacoustic fields within the system and thus provides valuable insight for optimizing its performance as a function of tunable parameters, such as particle radius, interfacial (Kapitza) resistance—modifiable through surface functionalization—, the fluid viscosity, and the frequency and intensity of the laser excitation. Finally, a refined analysis of the resonance frequencies emerging in the high-frequency regime is presented, and the effects of the fluid viscosity are examined.

## II. PROBLEM STATEMENT

We introduce here the general balance equations describing the coupling of the thermal and mechanical fields in fluid and solid phases, respectively. Then, we show how these equations are simplified for studying the time-harmonic regime in a spherical geometry.

Concerning the fluid medium, the set of equations takes into account the conservation of mass, momentum, and energy, which can be written as[99]

$$\frac{1}{B_0}\frac{\partial p}{\partial t} = \alpha_0 \frac{\partial T}{\partial t} - \vec{\nabla}\cdot\vec{v},$$
$$\rho_0 \frac{\partial \vec{v}}{\partial t} = -\vec{\nabla}p + \eta_0 \nabla^2 \vec{v} + (\xi_0 + \eta_0)\vec{\nabla}(\vec{\nabla}\cdot\vec{v}), \quad (1)$$
$$\rho_0 C_{p0} \frac{\partial T}{\partial t} = \kappa_0 \nabla^2 T + \alpha_0 T_e \frac{\partial p}{\partial t},$$

where the pressure $p$ [Pa], the temperature variation $T$ [K] and the particle velocity vector $\vec{v}$ [m/s] are the main variables depending on time $t$ [s] and space $\vec{r}$ [m]. Moreover, $\rho_0$ is the density [kg/m$^3$], $B_0$ the bulk modulus [Pa], $\alpha_0$ the coefficient of volumetric expansion [1/K], $\eta_0$ and $\xi_0$ the first and second viscosity coefficients [Pa s], $C_{p0}$ the specific heat at constant pressure [J/(kg K)], $T_e$ the ambient or equilibrium temperature [K] and, finally, $\kappa_0$ the thermal conductivity [W/(m K)]. Subscript 0 means that the parameters refer to the fluid phase outside the spherical particle. Similar parameters with subscript 1 will be introduced for the particle itself. It is important to remark that the balance equations given in Eq.(1) represent the combination of the linearized classical conservation laws with the linearized constitutive equations of the material. This linearization can be easily justified in our context since thermoacoustic waves are usually represented by small variations of the relevant quantities around given equilibrium values. As an example, $T$ is the variation of temperature with respect to its equilibrium value $T_e$ (the actual temperature being equal to $T + T_e$). This means that $|T| \ll T_e$ must always hold, and a similar relationship must be verified for each physical field.

We underline that in the absence of the assumption of small variations in the physical variables, the governing system becomes strongly nonlinear and, consequently, amenable only to numerical treatment. The objective of the present work is instead to derive an analytical solution which provides a tractable framework and preserves a clear interpretation of the underlying physical mechanisms. Concerning the admissible ranges of the physical variables, it is worth emphasizing that the fully coupled system, prior to linearization, is already nearly linear with respect to temperature. By contrast, the full mechanical model exhibits pronounced nonlinear behavior. As a result, after linearization, the regime of validity naturally allows for comparatively larger variations in temperature than in mechanical quantities such as forces and velocities, which must remain within a genuinely small-perturbation regime. Moreover, although all material parameters entering the model are, in principle, temperature-dependent, their variation remains moderate within the range relevant to practical applications. This justifies treating them as effectively constant to leading order, consistently with the adopted linear approximation.

A similar set of equations can be written for the solid medium by introducing the particle displacement vector $\vec{u}$ [m], the Lamé elastic coefficients $\lambda_1$ and $\mu_1$ [Pa], the specific heat at constant volume $C_{v1}$ [J/(kg K)], the externally applied body forces $\vec{b}_1$ [N] and the supplied thermal power density $Q_1$ [W/m$^3$]. The other parameters used for the solid have the same meaning as those introduced for fluids, with the only change being the index from 0 to 1. The classical continuum mechanics delivers[100]

$$\rho_1 \frac{\partial^2 \vec{u}}{\partial t^2} = (\lambda_1 + \mu_1)\vec{\nabla}(\vec{\nabla}\cdot\vec{u}) + \mu_1 \nabla^2 \vec{u} + \vec{b}_1$$
$$+ (\xi_1 + \eta_1)\vec{\nabla}(\vec{\nabla}\cdot\vec{v}) + \eta_1 \nabla^2 \vec{v} - \alpha_1 B_1 \vec{\nabla}T, \quad (2)$$
$$\rho_1 C_{v1}\frac{\partial T}{\partial t} = \kappa_1 \nabla^2 T - \alpha_1 B_1 \frac{\partial}{\partial t}\left(\vec{\nabla}\cdot\vec{u}\right) T_e + Q_1,$$

which is the system of equations governing the thermoelasticity under the hypotheses of small deformation $\hat{\varepsilon} = 1/2(\vec{\nabla}\vec{u} + \vec{\nabla}\vec{u}^T)$, and small temperature variations $T$ around $T_e$. While the first equation represents the momentum conservation, the second one describes the energy balance. We remark that in the solid medium we always have $\vec{v} = \partial \vec{u}/\partial t$, and $B_1 = \lambda_1 + (2/3)\mu_1$. Furthermore, we underline that the power density $Q_1$, entering the active solid medium, will represent the energy supplied to the system and converted into an acoustical wave through the photo-thermo-acoustic coupling. Typically, $Q_1$ will be generated by the photothermal effect, induced by a laser pulse. Used in different applications, not discussed in this context, we can also mention the Joule effect, induced by an electric current applied to the active material.

We finally remember that in both fluid and solid media the thermodynamic relation $\rho_i(C_{pi} - C_{vi}) = \alpha_i^2 B_i T_e$ ($i = 0, 1$) is always satisfied.

We now introduce the harmonic time-dependence through the substitution $\partial/\partial t \to i\omega$, and the spherical symmetry by means of the results discussed in Appendix A. For the fluid around the spherical particle, we obtain

$$\frac{i\omega}{B_0}p = \alpha_0 i\omega T - \frac{1}{r^2}\frac{d}{dr}\left(r^2 v\right), \quad (3)$$

$$i\omega\rho_0 v = -\frac{dp}{dr} + (\xi_0 + 2\eta_0)\frac{d}{dr}\left[\frac{1}{r^2}\frac{d}{dr}\left(r^2 v\right)\right], \quad (4)$$

$$i\omega\rho_0 C_{p0} T = \kappa_0 \frac{1}{r^2}\frac{d}{dr}\left(r^2 \frac{dT}{dr}\right) + i\omega\alpha_0 T_e p. \quad (5)$$

Similarly, for the spherical solid particle, we have under the same assumptions

$$-\omega^2 \rho_1 u = \left(\frac{\lambda_1 + 2\mu_1}{i\omega} + \xi_1 + 2\eta_1\right)\frac{d}{dr}\left[\frac{1}{r^2}\frac{d}{dr}\left(r^2 v\right)\right] - \alpha_1 B_1 \frac{dT}{dr}, \quad (6)$$

$$i\omega\rho_1 C_{v1} T = \kappa_1 \frac{1}{r^2}\frac{d}{dr}\left(r^2 \frac{dT}{dr}\right) - \alpha_1 B_1 i\omega T_e \frac{1}{r^2}\frac{d}{dr}\left(r^2 u\right) + Q_1, \quad (7)$$

where $v = i\omega u$, and we assumed $\vec{b}_1 = 0$, i.e. absence of body forces.

It is important to stress that any thermoacoustic field $\mathscr{F}(r,t)$ is finally obtained in the form $\mathscr{F}(r,t) = \mathscr{F}_e + \mathfrak{Re}\left[\mathscr{F}(r)e^{i\omega t}\right]$, where $\mathscr{F}_e$ is the equilibrium value used in the linearization of balance and constitutive equations, and $\mathscr{F}(r)$ is the complex solution obtained from Eqs.(3), (4), and (5), for the fluid and Eqs.(6), (7) for the solid particle. Of course, the field $\mathscr{F}$ corresponds to the temperature, the velocity, the normal stress or the heat flux.

## III. SOLUTION WITHIN THE EXTERNAL VISCOUS FLUID

We obtain here the general solution for the thermoacoustic fields in the fluid phase described by Eqs. (3), (4), and (5). From Eq. (3), we get

$$p = \alpha_0 B_0 T - \frac{B_0}{i\omega}\frac{1}{r^2}\frac{d}{dr}\left(r^2 v\right), \quad (8)$$

$$\frac{dp}{dr} = \alpha_0 B_0 \frac{dT}{dr} - \frac{B_0}{i\omega}\frac{d}{dr}\left[\frac{1}{r^2}\frac{d}{dr}\left(r^2 v\right)\right]. \quad (9)$$

Now, we substitute Eq.(8) in Eq.(5), and Eq.(9) in Eq.(4). In the relation obtained by substituting Eq.(8) in Eq.(5), we apply the thermodynamic relation $\rho_0(C_{p0} - C_{v0}) = \alpha_0^2 B_0 T_e$, and we obtain

$$i\omega\rho_0 C_{v0} T = \kappa_0 \frac{1}{r^2}\frac{d}{dr}\left(r^2 \frac{dT}{dr}\right) - \alpha_0 B_0 T_e \frac{1}{r^2}\frac{d}{dr}\left(r^2 v\right). \quad (10)$$

The relation obtained by inserting Eq.(9) in Eq.(4) can be simplified as follows

$$i\omega\rho_0 v = -\alpha_0 B_0 \frac{dT}{dr} + \left(\frac{B_0}{i\omega} + \xi_0 + 2\eta_0\right)\frac{d}{dr}\left[\frac{1}{r^2}\frac{d}{dr}\left(r^2 v\right)\right]. \quad (11)$$

At this point, we can determine the quantity $\frac{1}{r^2}\frac{d}{dr}\left(r^2 v\right)$ from Eq.(10), and we get

$$\frac{1}{r^2}\frac{d}{dr}\left(r^2 v\right) = -\frac{i\omega\rho_0 C_{v0}}{\alpha_0 B_0 T_e}T + \frac{\kappa_0}{\alpha_0 B_0 T_e}\frac{1}{r^2}\frac{d}{dr}\left(r^2 \frac{dT}{dr}\right). \quad (12)$$

Then, we substitute this result into Eq.(11), eventually obtaining the expression of the velocity in terms of the temperature only. We obtain

$$v = \mathscr{L}_0^\alpha \frac{dT}{dr} + \mathscr{L}_0^\beta \frac{d}{dr}\left(\mathbb{D}T\right), \quad (13)$$

where we introduced the coefficients

$$\mathscr{L}_0^\alpha = -\frac{\alpha_0 B_0}{i\omega\rho_0} - \left(\frac{B_0}{i\omega} + \xi_0 + 2\eta_0\right)\frac{C_{v0}}{\alpha_0 B_0 T_e}, \quad (14)$$

$$\mathscr{L}_0^\beta = \frac{1}{i\omega\rho_0}\left(\frac{B_0}{i\omega} + \xi_0 + 2\eta_0\right)\frac{\kappa_0}{\alpha_0 B_0 T_e}, \quad (15)$$

and the operator

$$\mathbb{D}f(r) = \frac{1}{r^2}\frac{d}{dr}\left(r^2 \frac{df}{dr}\right). \quad (16)$$

We can see from Eqs. (8) and (13) that if we know the temperature profile $T(r)$, then we can also calculate velocity and pressure functions $v(r)$ and $p(r)$. Hence, we look for a pure equation in the temperature. To do this, we apply the operator $\frac{1}{r^2}\frac{d}{dr}\left(r^2 \bullet\right)$ to Eq.(11), and then we substitute Eq.(12) in the resulting expression. We eventually obtain the following equation for the temperature behavior

$$\left(\frac{B_0}{i\omega} + \xi_0 + 2\eta_0\right)\kappa_0 \mathbb{D}^2 T$$
$$-\left[i\omega\rho_0 C_{v0}\left(\frac{B_0}{i\omega} + \xi_0 + 2\eta_0\right) + i\omega\rho_0\kappa_0 + \alpha_0^2 B_0^2 T_e\right]\mathbb{D}T$$
$$-\omega^2\rho_0^2 C_{v0} T = 0. \quad (17)$$

This equation can be written in the simplified form $a_0 \mathbb{D}^2 T + b_0 \mathbb{D} T + c_0 T = 0$, where $a_0$, $b_0$, and $c_0$ can be identified by means of Eq.(17). Interestingly, the same equation can be also rewritten as $(\mathbb{D} - \vartheta_{0T}^2)(\mathbb{D} - \vartheta_{0A}^2)T = 0$, where $\vartheta_{0T}$ and $\vartheta_{0A}$ are the solutions of the algebraic equation $a_0 \vartheta^2 + b_0 \vartheta + c_0 = 0$. The two subscripts $A$ and $T$ identify the acoustic and thermal modes of the process, respectively. This entails assuming $\vartheta_{0T} \neq \vartheta_{0A}$ on the basis of the typical physical parameters. We remark that the two operators $\mathbb{D} - \vartheta_{0T}^2$ and $\mathbb{D} - \vartheta_{0A}^2$ are commuting, as can be easily verified, and therefore the solutions of Eq.(17) are the linear combinations of the solutions of the two reduced equations $(\mathbb{D} - \vartheta_{0T}^2)T = 0$ and $(\mathbb{D} - \vartheta_{0A}^2)T = 0$.

Therefore, we search now the solution of the equation $(\mathbb{D} - \vartheta^2)T = 0$, for an arbitrary $\vartheta$, which corresponds to

$$\frac{1}{r^2}\frac{d}{dr}\left(r^2\frac{dT}{dr}\right) - \vartheta^2 T = 0, \quad (18)$$

or

$$\frac{d^2T}{dr^2} + \frac{2}{r}\frac{dT}{dr} - \vartheta^2 T = 0. \quad (19)$$

By using the substitution $T(r) = S(r)/r$, we obtain the following simple equation for $S(r)$

$$\frac{d^2 S}{dr^2} - \vartheta^2 S = 0, \quad (20)$$

with solutions

$$S(r) = e^{\pm \vartheta r}, \quad T(r) = \frac{1}{r} e^{\pm \vartheta r}. \quad (21)$$

Finally, the general solution for Eq.(17) is given by

$$T(r) = \frac{M_0}{r} e^{+\vartheta_{0T} r} + \frac{E_0}{r} e^{-\vartheta_{0T} r} + \frac{N_0}{r} e^{+\vartheta_{0A} r} + \frac{F_0}{r} e^{-\vartheta_{0A} r}, \quad (22)$$

where $\vartheta_{0T}$ and $\vartheta_{0A}$ are the roots of the associated second-degree equation, and we introduced the unknown coefficients $M_0$, $E_0$, $N_0$, and $F_0$. If we define the values of $\vartheta_{0T}$ and $\vartheta_{0A}$ such that $\mathrm{Re}(\vartheta_{0T}) < 0$, $\mathrm{Im}(\vartheta_{0T}) < 0$, and $\mathrm{Re}(\vartheta_{0A}) < 0$, $\mathrm{Im}(\vartheta_{0A}) < 0$, we will have $E_0 = 0$ and $F_0 = 0$. These signs derive from the following reasoning. First, it is assumed that there are only waves propagating outward from the spherical particle. Therefore, the waves must be progressive, i.e., they must propagate in the positive radial direction. The generic wave $e^{+\vartheta_{0T} r}$ propagates in the direction of positive $r$ only if the imaginary part of $\vartheta_{0T}$ is negative (please note that $e^{+\vartheta_{0T} r} e^{i\omega t} = e^{\mathrm{Re}(\vartheta_{0T})r + i\mathrm{Im}(\vartheta_{0T})r + i\omega t}$). Furthermore, dissipation (induced by viscosity and thermal conduction) must produce a decrease in the same direction, and therefore we must have a negative real part of $\vartheta_{0T}$. This is true for both thermal and acoustic components. In contrast, exponentials with a positive sign of real and imaginary parts are regressive waves that are not physically acceptable in our case, where the source is represented only by the central solid sphere. Hence,

$$T(r) = \frac{M_0}{r} e^{+\vartheta_{0T} r} + \frac{N_0}{r} e^{+\vartheta_{0A} r}. \quad (23)$$

It means that we will have to determine the two constants $M_0$ and $N_0$ to characterize the thermoacoustic response within the fluid.

We can use Eq.(13) to determine the velocity field. If we consider the general expression $T(r) = \frac{1}{r} e^{+\vartheta r}$, it can be easily proved that

$$\frac{dT}{dr} = \frac{1}{r^2} e^{+\vartheta r} (\vartheta r - 1), \quad (24)$$

$$\frac{d}{dr}(\mathbb{D}T) = \vartheta^2 \frac{dT}{dr} = \frac{\vartheta^2}{r^2} e^{+\vartheta r} (\vartheta r - 1). \quad (25)$$

Therefore, the velocity assumes the form

$$v(r) = \frac{M_0}{r^2} e^{+\vartheta_{0T} r} \left(\mathscr{L}_0^\alpha + \mathscr{L}_0^\beta \vartheta_{0T}^2\right)(\vartheta_{0T} r - 1)$$
$$+ \frac{N_0}{r^2} e^{+\vartheta_{0A} r} \left(\mathscr{L}_0^\alpha + \mathscr{L}_0^\beta \vartheta_{0A}^2\right)(\vartheta_{0A} r - 1). \quad (26)$$

We can also define the heat flux field through the relation $q(r) = -\kappa_0 \frac{dT}{dr}$. By using Eqs.(23) and (24), we obtain

$$q(r) = -\kappa_0 \frac{M_0}{r^2} e^{+\vartheta_{0T} r} (\vartheta_{0T} r - 1) \quad (27)$$

$$- \kappa_0 \frac{N_0}{r^2} e^{+\vartheta_{0A} r} (\vartheta_{0A} r - 1). \quad (28)$$

To conclude, we can find the complete expression for the pressure field. We use Eq.(8) combined with Eqs.(23) and (26), and we get

$$p(r) = \frac{M_0 B_0}{r} e^{+\vartheta_{0T} r} \left[\alpha_0 - \frac{1}{i\omega}\left(\mathscr{L}_0^\alpha + \mathscr{L}_0^\beta \vartheta_{0T}^2\right)\vartheta_{0T}^2\right]$$
$$+ \frac{N_0 B_0}{r} e^{+\vartheta_{0A} r} \left[\alpha_0 - \frac{1}{i\omega}\left(\mathscr{L}_0^\alpha + \mathscr{L}_0^\beta \vartheta_{0A}^2\right)\vartheta_{0A}^2\right]. \quad (29)$$

An equivalent simplified form can be obtained by substituting Eq. (12) into Eq. (3), and using again $\rho_0(C_{p0} - C_{v0}) = \alpha_0^2 B_0 T_e$, as follows

$$p(r) = \frac{M_0}{r} e^{+\vartheta_{0T} r} \frac{i\omega \rho_0 C_{p0} - \kappa_0 \vartheta_{0T}^2}{i\omega \alpha_0 T_E}$$
$$+ \frac{N_0}{r} e^{+\vartheta_{0A} r} \frac{i\omega \rho_0 C_{p0} - \kappa_0 \vartheta_{0A}^2}{i\omega \alpha_0 T_E}. \quad (30)$$

This quantity represents the pressure corresponding to the isotropic stress tensor $\hat{\sigma}_p = -p\hat{I}$. However, it is important to also consider the stress tensor $\hat{\sigma}_v$ related to the viscous response of the fluid, which is defined as

$$\hat{\sigma}_v = 2\eta_0 \hat{D} + \xi_0 \hat{I} \mathrm{tr}\hat{D}, \quad (31)$$

where $\hat{D} = \frac{1}{2}\left(\vec{\nabla}\vec{v} + \vec{\nabla}\vec{v}^T\right)$ is the rate of deformation (or strain rate tensor). In our geometry with spherical symmetry, the velocity field is given by $\vec{v} = \frac{\vec{r}}{r} v(r)$, and therefore, we calculate the derivative

$$\frac{\partial v_i}{\partial x_j} = \delta_{i,j}\frac{v}{r} + x_i x_j \frac{r\frac{dv}{dr} - v}{r^3}, \quad (32)$$

and therefore we get

$$D_{i,j} = \frac{1}{2}\left(\frac{\partial v_i}{\partial x_j} + \frac{\partial v_j}{\partial x_i}\right) = \delta_{i,j}\frac{v}{r} + x_i x_j \frac{r\frac{dv}{dr} - v}{r^3}. \quad (33)$$

From this expression, we easily obtain the trace of $\hat{D}$ as

$$\mathrm{tr}\hat{D} = \frac{dv}{dr} + 2\frac{v}{r} = \frac{1}{r^2}\frac{d}{dr}(r^2 v). \quad (34)$$

Now, the total stress tensor in the fluid is given by $\hat{\sigma}_{tot} = \hat{\sigma}_p + \hat{\sigma}_v$, and then the total radial traction is calculated as $\tilde{p} = \vec{n} \cdot \hat{\sigma}_{tot}\vec{n}$, where $\vec{n} = \vec{r}/r$. More explicitly, we have

$$\tilde{p}(r) = -p + (\xi_0 + 2\eta_0)\frac{dv}{dr} + 2\xi_0 \frac{v}{r}, \quad (35)$$

where we used the property $\vec{n} \cdot \hat{D}\vec{n} = D_{i,j}n_i n_j = \frac{dv}{dr}$. By using Eq.(8), and considering the second equality in Eq.(34), we obtain the traction expression

$$\tilde{p} = \left(\xi_0 + 2\eta_0 + \frac{B_0}{i\omega}\right)\frac{dv}{dr} + 2\left(\xi_0 + \frac{B_0}{i\omega}\right)\frac{v}{r} - \alpha_0 B_0 T, \quad (36)$$

which delivers the following final result

$$\begin{aligned}\tilde{p}(r) = & M_0 e^{+\vartheta_{0T}r}\left[-\frac{\alpha_0 B_0}{r}\right.\\
& + \frac{2}{r^3}\left(\xi_0 + \frac{B_0}{i\omega}\right)\left(\mathcal{L}_0^\alpha + \mathcal{L}_0^\beta \vartheta_{0T}^2\right)(\vartheta_{0T}r - 1)\\
& + \frac{1}{r^3}\left(\xi_0 + 2\eta_0 + \frac{B_0}{i\omega}\right)\left(\mathcal{L}_0^\alpha + \mathcal{L}_0^\beta \vartheta_{0T}^2\right)\\
& \left.\times (\vartheta_{0T}^2 r^2 - 2\vartheta_{0T}r + 2)\right]\\
& + N_0 e^{+\vartheta_{0A}r}\left[-\frac{\alpha_0 B_0}{r}\right.\\
& + \frac{2}{r^3}\left(\xi_0 + \frac{B_0}{i\omega}\right)\left(\mathcal{L}_0^\alpha + \mathcal{L}_0^\beta \vartheta_{0A}^2\right)(\vartheta_{0A}r - 1)\\
& + \frac{1}{r^3}\left(\xi_0 + 2\eta_0 + \frac{B_0}{i\omega}\right)\left(\mathcal{L}_0^\alpha + \mathcal{L}_0^\beta \vartheta_{0A}^2\right)\\
& \left.\times (\vartheta_{0A}^2 r^2 - 2\vartheta_{0A}r + 2)\right]. \quad (37)\end{aligned}$$

We will discuss the boundary conditions fulfilled by these fields after the study of the same quantities within the solid spherical particle.

## IV. SOLUTION WITHIN THE SPHERICAL PARTICLE

We search now for the solution of the thermoelastic fields within the spherical particle described by Eqs. (6) and (7). To begin, we can determine the quantity $\frac{1}{r^2}\frac{d}{dr}(r^2 v)$ from Eq.(7) (we remember that $v = i\omega u$), and we get

$$\frac{1}{r^2}\frac{d}{dr}(r^2 v) = \frac{\kappa_1}{\alpha_1 B_1 T_e}\mathbb{D}T - \frac{i\omega\rho_1 C_{v1}}{\alpha_1 B_1 T_e}T + \frac{Q_1}{\alpha_1 B_1 T_e}, \quad (38)$$

We suppose now that the heat source applied to the particle is uniformly distributed within the whole volume, and therefore we assume that $\frac{dQ_1}{dr} = 0$. We can now substitute Eq.(38) into Eq.(6), eventually obtaining

$$v = \mathcal{L}_1^\alpha \frac{dT}{dr} + \mathcal{L}_1^\beta \frac{d}{dr}(\mathbb{D}T), \quad (39)$$

where we introduced the coefficients

$$\mathcal{L}_1^\alpha = -\frac{\alpha_1 B_1}{i\omega\rho_1} - \left(\frac{\lambda_1 + 2\mu_1}{i\omega} + \xi_1 + 2\eta_1\right)\frac{C_{v1}}{\alpha_1 B_1 T_e}, \quad (40)$$

$$\mathcal{L}_1^\beta = \frac{1}{i\omega\rho_1}\left(\frac{\lambda_1 + 2\mu_1}{i\omega} + \xi_1 + 2\eta_1\right)\frac{\kappa_1}{\alpha_1 B_1 T_e}, \quad (41)$$

and the operator $\mathbb{D}T$ is defined in Eq.(16). To obtain a pure equation in the temperature field, we can combine Eq.(39) with Eq.(38), and we get

$$\begin{aligned}&\left(\frac{\lambda_1 + 2\mu_1}{i\omega} + \xi_1 + 2\eta_1\right)\kappa_1 \mathbb{D}^2 T\\
& - \left[i\omega\rho_1 C_{v1}\left(\frac{\lambda_1 + 2\mu_1}{i\omega} + \xi_1 + 2\eta_1\right) + i\omega\rho_1\kappa_1\right.\\
& \left.+ \alpha_1^2 B_1^2 T_e\right]\mathbb{D}T - \omega^2 \rho_1^2 C_{v1} T = i\omega\rho_1 Q_1, \quad (42)\end{aligned}$$

which is the solid counterpart of Eq.(17), previously obtained for the fluid medium. The only differences between the equations for solid and fluid are the presence of $\lambda_1 + 2\mu_1$ instead of $B_0$, and the presence of the term depending on the heat $Q_1$, entering the spherical particle. As before, the obtained equation for $T$ can be written in the simplified form $a_1 \mathbb{D}^2 T + b_1 \mathbb{D}T + c_1 T = 0$, where $a_1$, $b_1$, and $c_1$ can be identified by means of Eq.(42). Therefore, the equation can be also rewritten as $(\mathbb{D} - \vartheta_{1T}^2)(\mathbb{D} - \vartheta_{1A}^2)T = 0$, where $\vartheta_{1T}$ and $\vartheta_{1A}$ are the solutions of the algebraic equation $a_1 \vartheta^2 + b_1 \vartheta + c_1 = 0$. We remember that the two subscripts $A$ and $T$ identify the acoustic and thermal modes of the process, respectively.

We can follow the same procedure introduced for the fluid equation, and we obtain the following temperature solution for the solid sphere

$$\begin{aligned}T(r) = & \frac{V_1}{r}e^{+\vartheta_{1T}r} + \frac{E_1}{r}e^{-\vartheta_{1T}r}\\
& + \frac{W_1}{r}e^{+\vartheta_{1A}r} + \frac{F_1}{r}e^{-\vartheta_{1A}r} + \frac{Q_1}{i\omega\rho_1 C_{v1}}, \quad (43)\end{aligned}$$

where we introduced the coefficients $V_1$, $E_1$, $W_1$ and $F_1$. We remark that the particular solution (last term) has been obtained by considering the homogeneity of $Q_1$ within the sphere. We define the values of $\vartheta_{1T}$ and $\vartheta_{1A}$ such that $\mathrm{Re}(\vartheta_{1T}) < 0$, $\mathrm{Im}(\vartheta_{1T}) < 0$, and $\mathrm{Re}(\vartheta_{1A}) < 0$, $\mathrm{Im}(\vartheta_{1A}) < 0$. The negative sign of the real part, in conjunction with the negative sign of the imaginary part, reminds us that dissipation decreases the amplitude of the wave in the direction of propagation. Within the particle, we can consider both progressive and regressive radial waves (characterized by the positive and negative sign in the exponent, respectively) as the interface between solid and fluid allows for thermal and mechanical reflections. This means that we can keep all four terms in the formula, two relating to the acoustic component and two relating to the thermal component. By means of Eq.(39), we can find the corresponding solution for the velocity field. Straightforward calculations based on Eqs. (24) and (25) deliver the expression

$$\begin{aligned}v(r) = & \left(\mathcal{L}_1^\alpha + \mathcal{L}_1^\beta \vartheta_{1T}^2\right)\\
& \times \frac{V_1 e^{+\vartheta_{1T}r}(\vartheta_{1T}r - 1) - E_1 e^{-\vartheta_{1T}r}(\vartheta_{1T}r + 1)}{r^2}\\
& + \left(\mathcal{L}_1^\alpha + \mathcal{L}_1^\beta \vartheta_{1A}^2\right)\\
& \times \frac{W_1 e^{+\vartheta_{1A}r}(\vartheta_{1A}r - 1) - F_1 e^{-\vartheta_{1A}r}(\vartheta_{1A}r + 1)}{r^2}. \quad (44)\end{aligned}$$

We remark that the entering heat $Q_1$ does not directly affect the velocity field since it is uniform, and $v$ depends directly on the derivative $dT/dr$. Within the particle, we need to avoid singularities in both $T(r)$ and $v(r)$ for $r \to 0$, and then there are some conditions to be imposed on the introduced coefficients. Since for small values of $x$ we have $e^x \sim 1 + x$, from Eq.(43) we obtain the first order development

$$T(r) \sim \frac{V_1 + E_1 + W_1 + F_1}{r} + (V_1 - E_1)\vartheta_{1T} \\ + (W_1 - F_1)\vartheta_{1A} + \frac{Q_1}{i\omega\rho_1 C_{v1}}, \quad (45)$$

which is valid for small values of $r$. Therefore, we have to impose the relationship $V_1 + E_1 + W_1 + F_1 = 0$ to eliminate the singularity for $r \to 0$ of type $1/r$. We can perform a similar analysis for the velocity field. In this case, the first-order development for small values of $r$ is given by

$$v(r) \sim \left(\mathcal{L}_1^\alpha + \mathcal{L}_1^\beta \vartheta_{1T}^2\right)(V_1 + E_1)\frac{\vartheta_{1T}^2 r^2/2 - 1}{r^2} \\ + \left(\mathcal{L}_1^\alpha + \mathcal{L}_1^\beta \vartheta_{1A}^2\right)(W_1 + F_1)\frac{\vartheta_{1A}^2 r^2/2 - 1}{r^2}. \quad (46)$$

Therefore, to eliminate the singularity of type $1/r^2$ we have to impose $\left(\mathcal{L}_1^A + \mathcal{L}_1^B \vartheta_{1T}^2\right)(V_1 + E_1) + \left(\mathcal{L}_1^A + \mathcal{L}_1^B \vartheta_{1A}^2\right)(W_1 + F_1) = 0$. Combining the conditions obtained for the temperature and velocity fields, assuming that $\vartheta_{1T} \neq \vartheta_{1A}$, we deduce that we must satisfy the relations

$$V_1 + E_1 = 0, \quad (47)$$
$$W_1 + F_1 = 0. \quad (48)$$

These conditions help us to simplify the form of Eqs.(43) and (44). We introduce $M_1 = 2V_1$ and $N_1 = 2W_1$, finally obtaining

$$T(r) = M_1 \frac{\sinh(\vartheta_{1T}r)}{r} + N_1 \frac{\sinh(\vartheta_{1A}r)}{r} \\ + \frac{Q_1}{i\omega\rho_1 C_{v1}}, \quad (49)$$

$$v(r) = M_1 \left(\mathcal{L}_1^\alpha + \mathcal{L}_1^\beta \vartheta_{1T}^2\right) \\ \times \frac{\vartheta_{1T} r \cosh(\vartheta_{1T}r) - \sinh(\vartheta_{1T}r)}{r^2} \\ + N_1 \left(\mathcal{L}_1^\alpha + \mathcal{L}_1^\beta \vartheta_{1A}^2\right) \\ \times \frac{\vartheta_{1A} r \cosh(\vartheta_{1A}r) - \sinh(\vartheta_{1A}r)}{r^2}. \quad (50)$$

In addition to these fields, we can introduce the heat flow and the radial traction, which are useful for imposing the boundary conditions later. The heat flux is introduced through the relation $q(r) = -\kappa_1 \frac{dT}{dr}$, and therefore, by using Eq.(49), we obtain

$$q(r) = -M_1 \kappa_1 \frac{\vartheta_{1T} r \cosh(\vartheta_{1T}r) - \sinh(\vartheta_{1T}r)}{r^2} \\ - N_1 \kappa_1 \frac{\vartheta_{1A} r \cosh(\vartheta_{1A}r) - \sinh(\vartheta_{1A}r)}{r^2}. \quad (51)$$

Concerning the radial traction, we start by considering the total stress tensor $\hat{\sigma}_{tot}$ inside the spherical particle, given by

$$\hat{\sigma}_{tot} = 2\mu_1 \hat{\varepsilon} + \lambda_1 \hat{I} \mathrm{tr}\hat{\varepsilon} - \alpha_1 \left(\lambda_1 + \frac{2}{3}\mu_1\right) \hat{I} T \\ + i\omega \left(2\eta_1 \hat{\varepsilon} + \xi_1 \hat{I} \mathrm{tr}\hat{\varepsilon}\right). \quad (52)$$

It is composed of elastic, thermal, and viscous contributions, mentioned in order of appearance. Here $\hat{\varepsilon} = \frac{1}{2}\left(\vec{\nabla}\vec{u} + \vec{\nabla}\vec{u}^T\right)$ is the infinitesimal strain tensor measuring the solid deformation. As discussed in the case of the fluid medium, we are interested in the total radial traction calculated as $\tilde{p} = \vec{n} \cdot \hat{\sigma}_{tot} \vec{n}$, where $\vec{n} = \vec{r}/r$. Since

$$\varepsilon_{i,j} = \frac{1}{2}\left(\frac{\partial u_i}{\partial u_j} + \frac{\partial u_j}{\partial x_i}\right) = \delta_{i,j}\frac{u}{r} + x_i x_j \frac{r\frac{du}{dr} - u}{r^3}, \quad (53)$$

we have that

$$\mathrm{tr}\hat{\varepsilon} = \varepsilon_{k,k} = \frac{du}{dr} + 2\frac{u}{r}, \quad (54)$$

and $\varepsilon_{ij} n_i n_j = \frac{du}{dr}$. Summing up, we get the following expression for the radial traction

$$\tilde{p} = \left(\xi_1 + 2\eta_1 + \frac{\lambda_1 + 2\mu_1}{i\omega}\right)\frac{dv}{dr} + 2\left(\xi_1 + \frac{\lambda_1}{i\omega}\right)\frac{v}{r} \\ - \alpha_1 \left(\lambda_1 + \frac{2}{3}\mu_1\right) T, \quad (55)$$

where we introduced the velocity $v = i\omega u$. By using Eqs. (49) and (50), the previous relation assumes the following explicit form

$$\tilde{p}(r) = M_1 \left[-\alpha_1 B_1 \frac{\sinh(\vartheta_{1T}r)}{r} \\ + 2\left(\xi_1 + \frac{\lambda_1}{i\omega}\right)\left(\mathcal{L}_1^\alpha + \mathcal{L}_1^\beta \vartheta_{1T}^2\right) \\ \times \frac{\vartheta_{1T} r \cosh(\vartheta_{1T}r) - \sinh(\vartheta_{1T}r)}{r^3} \\ + \left(\xi_1 + 2\eta_1 + \frac{\lambda_1 + 2\mu_1}{i\omega}\right)\left(\mathcal{L}_1^\alpha + \mathcal{L}_1^\beta \vartheta_{1T}^2\right) \\ \times \frac{(\vartheta_{1T}^2 r^2 + 2)\sinh(\vartheta_{1T}r) - 2\vartheta_{1T} r \cosh(\vartheta_{1T}r)}{r^3}\right] \\ + N_1 \left[-\alpha_1 B_1 \frac{\sinh(\vartheta_{1A}r)}{r} \\ + 2\left(\xi_1 + \frac{\lambda_1}{i\omega}\right)\left(\mathcal{L}_1^\alpha + \mathcal{L}_1^\beta \vartheta_{1A}^2\right) \\ \times \frac{\vartheta_{1A} r \cosh(\vartheta_{1A}r) - \sinh(\vartheta_{1A}r)}{r^3} \\ + \left(\xi_1 + 2\eta_1 + \frac{\lambda_1 + 2\mu_1}{i\omega}\right)\left(\mathcal{L}_1^\alpha + \mathcal{L}_1^\beta \vartheta_{1A}^2\right) \\ \times \frac{(\vartheta_{1A}^2 r^2 + 2)\sinh(\vartheta_{1A}r) - 2\vartheta_{1A} r \cosh(\vartheta_{1A}r)}{r^3}\right] \\ - \alpha_1 B_1 \frac{Q_1}{i\omega\rho_1 C_{v1}}. \quad (56)$$

These results will be used for imposing the pertinent boundary conditions.

## V. SOLUTION OF THE COUPLED THERMOACOUSTIC PROBLEM

We have determined, in previous Sections, the general solutions for the thermoacoustic fields in each phase. Now, the general solution for the whole system can be obtained by imposing the interface condition at the fluid-solid interface for $r = R$.

These conditions represent the continuity of heat flux $q$, velocity $v$, and radial traction $\tilde{p}$: $[\![q]\!] = 0$, $[\![v]\!] = 0$, and $[\![\tilde{p}]\!] = 0$. Moreover, we have to impose the interface relation for the temperature jump $[\![T]\!]$, controlled by the Kapitza resistance,[130] denoted by $\tau_K$ [m$^2$K/W], as follows[101–103]

$$[\![T]\!] = T(R^+) - T(R^-) = -\tau_K q(R), \quad (57)$$

where

$$q(R) = -\kappa_1 \frac{dT(R^-)}{dr} = -\kappa_0 \frac{dT(R^+)}{dr}. \quad (58)$$

These four interface conditions allow for the computation of the four unknown coefficients $M_0$, $N_0$, $M_1$, and $N_1$. They can indeed be recast into the inhomogeneous linear system

$$\mathbf{A} \begin{pmatrix} M_0 \\ N_0 \\ M_1 \\ N_1 \end{pmatrix} = \begin{pmatrix} \dfrac{Q_1}{i\omega\rho_1 C_{v1}} \\ 0 \\ 0 \\ -\dfrac{\alpha_1 B_1 Q_1}{i\omega\rho_1 C_{v1}} \end{pmatrix}, \quad (59)$$

with the matrix $\mathbf{A}$ defined as

$$\mathbf{A} = \begin{pmatrix} s_0(\vartheta_{0T}) & s_0(\vartheta_{0A}) & -\ell_1(\vartheta_{1T}) & -\ell_1(\vartheta_{1A}) \\ g_0(\vartheta_{0T}) & g_0(\vartheta_{0A}) & -g_1(\vartheta_{1T}) & -g_1(\vartheta_{1A}) \\ h_0(\vartheta_{0T}) & h_0(\vartheta_{0A}) & -h_1(\vartheta_{1T}) & -h_1(\vartheta_{1A}) \\ f_0(\vartheta_{0T}) & f_0(\vartheta_{0A}) & -f_1(\vartheta_{1T}) & -f_1(\vartheta_{1A}) \end{pmatrix}, \quad (60)$$

where to compact the notation we introduced the following functions

$$s_0(\vartheta) = \ell_0(\vartheta) + \tau_K g_0(\vartheta), \quad (61)$$

$$\ell_0(\vartheta) = \frac{e^{+\vartheta r}}{r}, \quad g_0(\vartheta) = -\kappa_0 \frac{e^{+\vartheta r}}{r^2}(\vartheta r - 1), \quad (62)$$

$$h_0(\vartheta) = \left(\mathscr{L}_0^\alpha + \mathscr{L}_0^\beta \vartheta^2\right) \frac{e^{+\vartheta r}}{r^2}(\vartheta r - 1), \quad (63)$$

$$\begin{aligned}
f_0(\vartheta) = & e^{+\vartheta r} \left[ -\frac{\alpha_0 B_0}{r} + \frac{1}{r^3}\left(\xi_0 + 2\eta_0 + \frac{B_0}{i\omega}\right) \right. \\
& \times \left(\mathscr{L}_0^\alpha + \mathscr{L}_0^\beta \vartheta^2\right)(\vartheta^2 r^2 - 2\vartheta r + 2) \\
& \left. + \frac{2}{r^3}\left(\xi_0 + \frac{B_0}{i\omega}\right)\left(\mathscr{L}_0^\alpha + \mathscr{L}_0^\beta \vartheta^2\right)(\vartheta r - 1) \right],
\end{aligned} \quad (64)$$

for the fluid phase, and

$$\ell_1(\vartheta) = \frac{\sinh(\vartheta r)}{r}, \quad (65)$$

$$g_1(\vartheta) = -\kappa_1 \frac{\vartheta r \cosh(\vartheta r) - \sinh(\vartheta r)}{r^2}, \quad (66)$$

$$h_1(\vartheta) = \left(\mathscr{L}_1^\alpha + \mathscr{L}_1^\beta \vartheta^2\right) \frac{\vartheta r \cosh(\vartheta r) - \sinh(\vartheta r)}{r^2}, \quad (67)$$

$$\begin{aligned}
f_1(\vartheta) = & \left[ -\alpha_1 B_1 \frac{\sinh(\vartheta r)}{r} + 2\left(\xi_1 + \frac{\lambda_1}{i\omega}\right) \right. \\
& \times \left(\mathscr{L}_1^\alpha + \mathscr{L}_1^\beta \vartheta^2\right) \frac{\vartheta r \cosh(\vartheta r) - \sinh(\vartheta r)}{r^3} \\
& + \left(\xi_1 + 2\eta_1 + \frac{\lambda_1 + 2\mu_1}{i\omega}\right)\left(\mathscr{L}_1^\alpha + \mathscr{L}_1^\beta \vartheta^2\right) \\
& \left. \times \frac{(\vartheta^2 r^2 + 2)\sinh(\vartheta r) - 2\vartheta r \cosh(\vartheta r)}{r^3} \right],
\end{aligned} \quad (68)$$

for the solid phase. The linear system of equations defined in Eq. (59) can obviously be solved numerically very easily using classic linear algebra procedures for each frequency value.

## VI. THERMOACOUSTIC BEHAVIOR OF GOLD PARTICLES IN WATER

In this Section, we build upon the results obtained previously, which we briefly recapitulate here. In particular, we have derived the exact solution for the four fundamental physical fields—velocity, force, temperature, and heat flux—in a spherical geometry, both for the solid particle and for the surrounding fluid. These solutions are now coupled through the boundary conditions in order to investigate the dynamics of a spherical particle immersed in a fluid. The two analytical solutions previously obtained allow us to treat separately the two limiting configurations: (i) vanishing thermal expansion in the fluid ($\alpha_0 = 0$), and (ii) vanishing thermal expansion in the solid ($\alpha_1 = 0$). These limiting cases provide a natural decomposition of the fully coupled thermo-mechanical problem. It is important to recall that the general solution for each physical field is constructed as the superposition of these two particular solutions. This superposition principle follows from the linear structure of the governing equations under the adopted approximation scheme and enables a transparent interpretation of the respective contributions of solid and fluid thermal expansion.

We consider the thermoacoustic fields generated by a uniform time-harmonic thermal source (laser) applied to a spherical golden particle embedded in a water fluid matrix (see Fig. 1, right panel). For gold and water, we adopted the parameters reported in Table I, obtained from the literature.[68,107–111] In addition, we have considered a spherical particle with radius in the range $R = 50 \div 300$ nm, and an interface thermal resistance in the range $\tau_K = 10^{-9} \div 10^{-6}$ m$^2$ K/W.[101–103]

The parameters for the gold particle reported in Table I refer to bulk gold. It is important to note that these parameters may undergo changes at the nanometric level due to scale effects. For example, the Young's modulus is typically observed to decrease as particle size is reduced while thermal expansion coefficient increases with decrease in the size of nanomaterials[112–114]. However, these behaviors vary greatly from case to case, and therefore there are no definitive laws that are universally accepted[115]. Typically, these variations are only relevant at sizes smaller than a few tens of nm and can therefore be neglected, at least as a first approximation, for particles of biomedical interest. Some experiments on nanoparticle elastic vibrations have even highlighted the absence of scale effects down to the nanometer scale.[116,117]

A comment is warranted regarding the assumption of a *homogeneous* power density $Q_1$ when the heat is delivered to the nanoparticle by a laser beam. This assumption may appear questionable for nanoparticles whose linear dimensions significantly exceed the optical penetration depth within the material (on the order of a few tens of nanometers for visible wavelengths). However, for nanoparticles sizes of practical relevance, this issue can be neglected, since both diffusive and hot electrons[118–120] contribute to homogenizing the absorbed power density $Q_1$ throughout the particle on a time scale shorter than the mechanical and thermal dynamical time scales associated with the generation of the pressure wave in water.

In Fig. 2, we represent the generated acoustic pressure as a function of the frequency for four different values of the Kapitza resistances $\tau_K = 10^{-9}, 10^{-8}, 10^{-7}, 10^{-6}$ m$^2$ K/W (panels a, b, c, and d). In each plot, we have represented three different curves defined as follows. The black curves represent the total pressure generated within the system. The red curves represent the pressure generated by the thermophone mechanism, obtained by setting to zero the thermal expansion coefficient of the solid ($\alpha_1 = 0$). So doing, the acoustic waves in the fluid come solely from thermal compression and expansion within the fluid itself. The blue curves represent the pressure generated by the mechanophone mechanism. In this case, the thermal expansion coefficient of the fluid is set to zero ($\alpha_0 = 0$), allowing the generation of acoustic waves only through the size modulation of the particle. Of course, the sum of each red curve and blue curve gives the black curve as a result. Importantly, we observe that the Kapitza resistance is able to shift the crossover between thermophone and mechanophone mechanisms. The termophone behavior (red lines) is the dominant term at low frequencies, whereas the mechanophone (blue lines) is dominant at higher frequencies. Indeed, the heat flow decreases with increasing Kapitza resistance.

The Kapitza's resistance can be modified by appropriately functionalising the particle surface.[103,121,122] Specifically, increasing the Kapitza resistance reduces the heat flux transferred to the fluid, thereby limiting the temperature rise in the surrounding medium, while the mechanophone contribution—being primarily driven by elastic coupling—remains operative. This configuration may be advantageous in diagnostic applications, where minimizing fluid heating is de-sirable. Conversely, in therapeutic applications where local temperature elevation is required, lower interfacial resistances promote stronger thermal transfer to the fluid and are therefore acceptable or even preferable.

TABLE I. Physical properties of water and gold, used in the thermoacoustic model (measured at room temperature).[68,107–111]

| Water | Gold |
|---|---|
| $\rho_0 = 10^3$ Kg/m$^3$ | $\rho_1 = 19.32 \times 10^3$ Kg/m$^3$ |
| $\alpha_0 = 3.03 \times 10^{-4}$ 1/K | $\alpha_1 = 0.42 \times 10^{-4}$ 1/K |
| $C_{p0} = 4400$ J/(Kg K) | $C_{p1} = 130$ J/(Kg K) |
| $\kappa_0 = 0.607$ W/(m K) | $\kappa_1 = 310$ W/(m K) |
| $\eta_0 = 0.894 \times 10^{-3}$ Pa s | $\eta_1 \simeq 0$ Pa s |
| $\xi_0 = 1.5 \times 10^{-3}$ Pa s | $\xi_1 \simeq 0$ Pa s |
| $B_0 = 2.15 \times 10^9$ Pa | $B_1 = 178 \times 10^9$ Pa |
| $\mu_0 \simeq 0$ Pa | $\mu_1 = 27 \times 10^9$ Pa |
| $\lambda_0 = 2.15 \times 10^9$ Pa | $\lambda_1 = 160 \times 10^9$ Pa |

To better explain this shift, we observe that the termophone mechanism becomes negligible at high frequencies since the penetration length of the thermal wave in the fluid follows the approximated expression $\mathscr{L}_{th} = \sqrt{2\kappa_0/(\omega\rho_0 C_{p0})}$, decreasing as $1/\sqrt{\omega}$. The region identified by $R < r < R + n\mathscr{L}_{th}$ (with $n = 3$ or 5 depending on the considered approximation) is the active domain, where the thermophone effect generates the acoustic wave. Therefore, for high frequencies, $\mathscr{L}_{th}$ becomes very small, and the thermophone mechanism is strongly reduced. In this high-frequency region, the mechanophone mechanism (based on a piston effect, or dilation and compression of the solid particle) becomes dominant. Note that at extremely high frequencies, even the mechanophone mechanism is reduced, and this is due to the viscosity of the fluid, as described below. Please note that the fields represented in Fig. 2 are measured at the distance $R + 5\mathscr{L}_{th}$, where $\omega = 10$ MHz (the smallest frequency adopted, i.e., the largest $\mathscr{L}_{th}$), and therefore outside the active region in any case. This means that the pressure values represented are significant from an acoustic point of view because they are measured far from the region of generation.

For completeness, Fig. 3 shows the three-dimensional diagram of pressure as a function of frequency and measurement position of the acoustic field. It can be seen that as the frequency increases, the fields are increasingly localized near the fluid-solid interface, due to the dissipation phenomena described below.

In Figs. 2 and 3, we also observe that for high frequencies the pressure-frequency response exhibits resonance peaks (corresponding to the reflections of the elastic wave inside the particle). Therefore, we can identify a sub-resonance and a resonance regime.

It is worth noting that small peaks are also observed in the purely thermophone response (red curves in Fig. 2). In order to isolate this contribution, we set the thermal expansion coefficient of the solid particle equal to zero. Under this assumption, the particle does not undergo thermally induced volumetric deformation, while remaining fully elastically deformable. Although thermal expansion within the particle is suppressed, thermoacoustic waves are still generated in the surrounding fluid. These elastic waves exert a mechanical ac-tion on the particle surface, thereby inducing spherical elastic waves within the solid. When the frequency of the induced elastic waves approaches the characteristic resonance frequencies of the particle, distinct peaks appear in the response. Depending on the specific set of physical parameters, these features may correspond either to resonances or to anti-resonances. Indeed, parameters modify the fluid-solid coupling, so that the same mode may appear either as an enhanced (resonance) or a suppressed (anti-resonance) response.

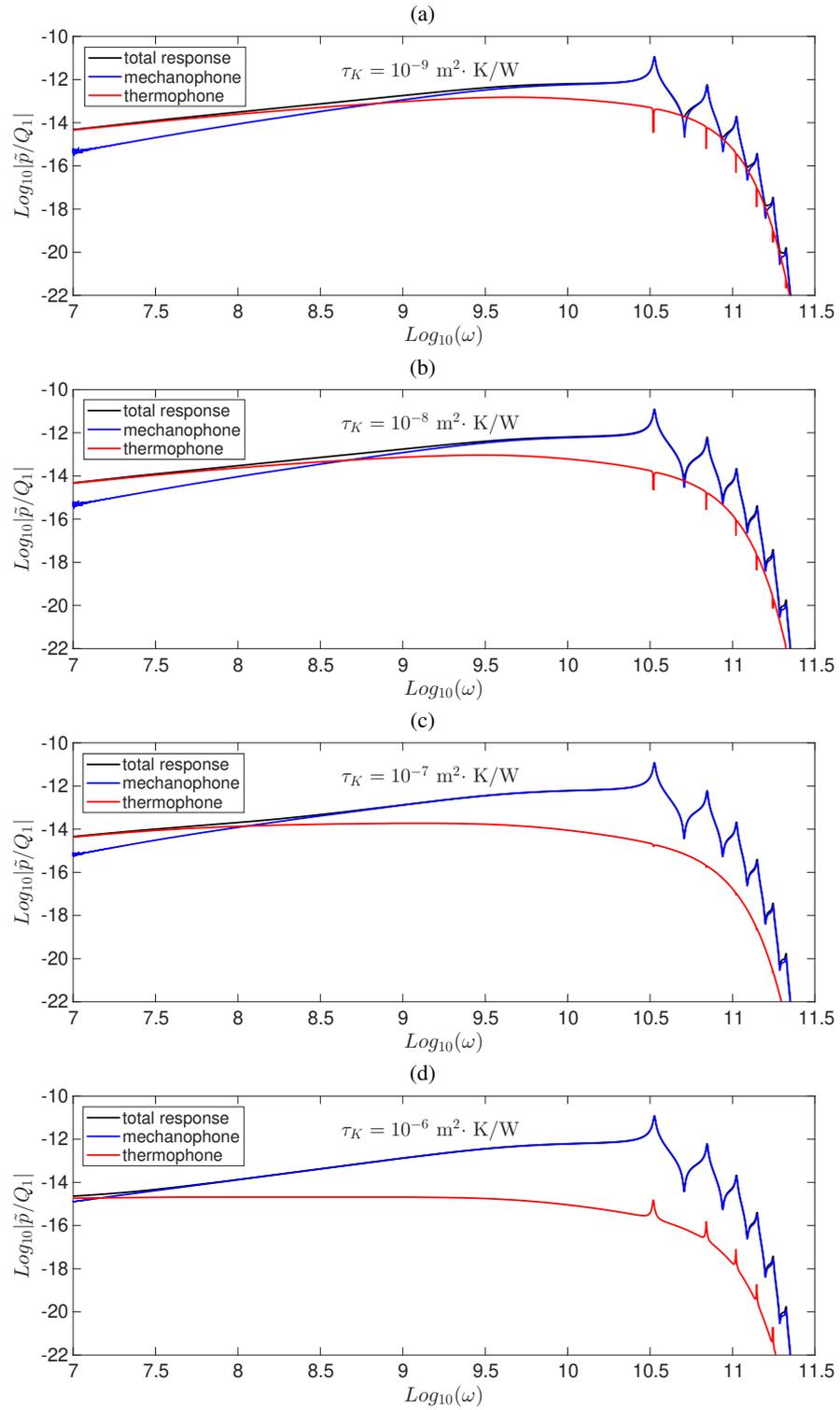

FIG. 2. Pressure-frequency response for the gold-water system with different values of the Kapitza resistance between the two phases. The four plots exhibit the shift of the crossover between thermophone and mechanophone behavior. The interfacial thermal resistance does not affect the resonance frequencies (moreover, the thermophone resonances change the sign with different values of $\tau_K$). We adopted the radius $R = 300$ nm and the Kapitza resistances $\tau_K = 10^{-9}, 10^{-8}, 10^{-7}, 10^{-6}$ m$^2$ K/W from the top to the bottom. The fields are measured at the distance $R + 5\mathscr{L}_{th}$, where $\mathscr{L}_{th} = \sqrt{2\kappa_0/(\omega \rho_0 C_{p0})} = 0.16 \mu$m is the thermal length at $\omega = 10$ MHz (the smallest frequency adopted).

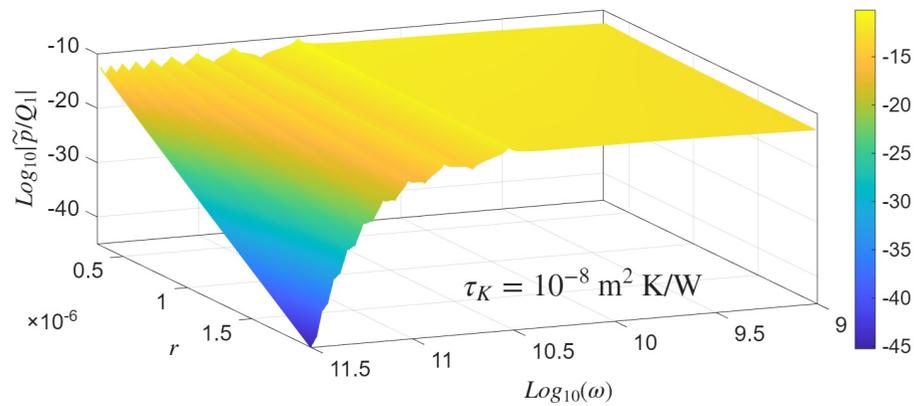

FIG. 3. Total pressure as a function of frequency $\omega$ and measurement position $r$. We adopted the parameters of gold and water with a thermal resistance $\tau_K = 10^{-8}$ m$^2$ K/W. Moreover, we considered a particle of radius $R = 300$ nm, and $R < r < 2\mu$m. Please note that the frequency axis is inverted with respect to Fig. 2 to improve the graphical representation.

The resonance peaks depend on the size of the spherical particle. Indeed, to better understand this resonance behavior, in Fig. 4 we plot the total pressure (thermophone plus mechanophone) versus the frequency for different particle radii $R = 50, 75, 100, 150, 225, 300$ nm, and different values of the Kapitza resistances $\tau_K = 10^{-9}, 10^{-7}$ m$^2$ K/W (panels a and b). We observe effectively that the resonance frequencies are dependent on the particle radius. In particular, we see that the first frequency resonance (highlighted by a dot) increases with a decreasing radius. This behavior is described by the general relation $f_{1,res} = C/(2\pi R)\sqrt{(\lambda_1 + 2\mu_1)/\rho_1}$, where $C$ is a constant parameter.[123,124] It is straightforward to verify that the represented dots follow this $1/R$ dependence exactly.

Furthermore, in Fig. 4, we observe that the pressure intensity at the first resonance peak decreases as the radius decreases. Consequently, it is interesting to note that there is a particle radius below which this peak no longer represents the global maximum of the pressure curve. We note that a radius of 300 nm has been selected for most of the plots in order to illustrate more clearly the behavior in the resonance region. In many practical applications, the particle size is smaller—typically on the order of a few tens of nanometers—but in such cases it becomes significantly more challenging to disentangle and discuss all aspects of the system's physics, particularly at high frequencies. It should also be recalled that the behavior of water in the ultrahigh-frequency regime is extremely complex and not yet fully understood.[125–129] For this reason, we have represented the frequency diagrams with a maximum frequency of around $10^{12}$ rad/sec.

To further understand the thermo-acoustic behavior of the particle, we represent in Fig. 5 some acoustic parameters that describe the propagation inside and outside the particle itself. In panel (a), we show the acoustic $\mathscr{L}_{ac} = -1/\Re e\,[\vartheta_{0A}]$ and thermal $\mathscr{L}_{th} = -1/\Re e\,[\vartheta_{0T}]$ penetration lengths in the fluid phase. In panel (b), we show the acoustic $\lambda_{ac} = -2\pi/\Im m\,[\vartheta_{1A}]$ and thermal $\lambda_{th} = -2\pi/\Im m\,[\vartheta_{1T}]$ wavelengths within the solid particle. All these parameters are represented versus the frequency $\omega$. We also added a dashed line corresponding to the radius $R = 300$ nm, for the sake of comparison. We recall here the approximate expression for the acoustic characteristic length[63]

$$\mathscr{L}_{ac} = \frac{2C_0^3}{\omega^2} \frac{1}{\frac{\xi_0 + 2\eta_0}{\rho_0} + \left(\frac{C_{p0}}{C_{v0}} - 1\right)\frac{\kappa_0}{\rho_0 C_{p0}}}, \quad (69)$$

where $C_0 = \sqrt{(B_0/\rho_0)(C_{p0}/C_{v0})}$ is the sound speed in the fluid phase, and the previously mentioned approximated expression $\mathscr{L}_{th} = \sqrt{2\kappa_0/(\omega \rho_0 C_{p0})}$ for the thermal characteristic length. In Fig. 5 (panel a), we can see that the $\mathscr{L}_{ac}$ curve follows the $1/\omega^2$ trend, and the $\mathscr{L}_{th}$ curve follows the $1/\sqrt{\omega}$ trend, in agreement with previous expressions. The physics of these curves can be summarized as follows. The penetration depth of the acoustic wave $\mathscr{L}_{ac}$ describes the dissipation phenomena of the process (thermal conduction and fluid viscosity) and therefore the decrease of the acoustic wave as it progresses through the fluid medium. It can be seen that this decrease is practically negligible at low frequencies but becomes very important at high frequencies (trend $1/\omega^2$), where the propagation length becomes comparable to the particle radius. We have already discussed the thermal penetration length $\mathscr{L}_{th}$, and observe here that it is always less than the particle radius (in the gold-water system). Furthermore, we can see that this parameter decreases with frequency (trend $1/\sqrt{\omega}$), driving the transition between thermophone and mechanophone mechanisms.

Concerning panel (b) of Fig. 5, a simple approximation leads to the expressions $\lambda_{th} = 2\pi\sqrt{2\kappa_1/(\omega \rho_1 C_{p1})}$, and $\lambda_{ac} = 2\pi/\omega \sqrt{B_1/\rho_1}$, whose frequency trends ($1/\sqrt{\omega}$, and $1/\omega$, respectively) correspond perfectly to the curves shown. The thermal wavelength $\lambda_{th}$ is much larger than the particle radius at low frequencies and becomes comparable to it only in the high-frequency regime. Consequently, at low frequencies the temperature and the associated heat flux remain nearly spatially uniform within the particle. As the frequency increases and $\lambda_{th}$ approaches the particle size, increasingly pronounced spatial gradients develop inside the particle. A similar scaling behavior characterizes the acoustic wavelength $\lambda_{ac}$, which likewise decreases with increasing frequency. When it becomes comparable to the particle radius, standing-wave conditions can be established, thereby accounting for the appearance of high-frequency resonance peaks.

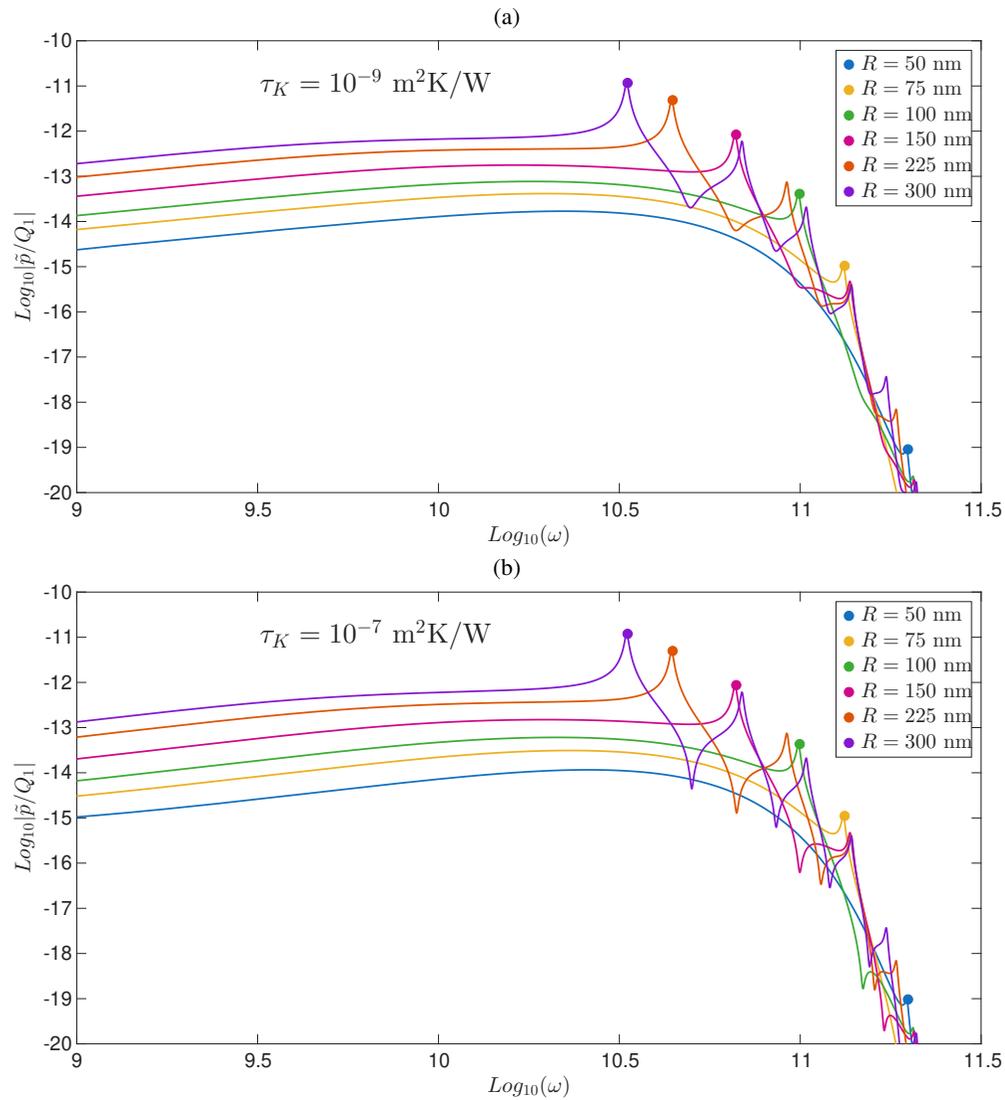

FIG. 4. Pressure-frequency response for the gold-water system with different particle sizes and values of the Kapitza resistance between the two phases. A larger radius gives larger resonance peaks at lower frequencies. In addition, there is a critical radius below which the first resonance peak is no longer the absolute maximum. We adopted the radius values $R$=50 (blue), $R$=75 (yellow), $R$=100 (green), 150 (light purple), 225 (red), 300 (dark purple) nm, and the Kapitza resistances $\tau_K = 10^{-9}, 10^{-7}$ m$^2$ K/W from the top to the bottom. The fields are measured at the distance $R + 5\mathscr{L}_{th}$, where $\mathscr{L}_{th} = \sqrt{2\kappa_0/(\omega \rho_0 C_{p0})} = 0.16 \mu$m is the thermal length at $\omega = 10$ MHz (the smallest frequency adopted).

These concepts can be easily understood by observing the variation in thermoacoustic fields as the radius varies for a fixed frequency. In Fig. 6, we find the evolution of the four main thermoacoustic fields (temperature, thermal flux, velocity, and pressure) versus the radius of the system in the sub-resonance frequency region. Similarly, in Fig. 7, we show the fields versus the radius of the system in the resonance frequency region. In Fig. 6, we adopted the frequencies $\omega = 10^7, 10^8, 10^9, 10^{10}$ rad/sec, and in Fig. 7, the values $\omega = 5 \times 10^{10}, 10 \times 10^{10}, 15 \times 10^{10}, 20 \times 10^{10}$ rad/sec. In both figures, we have used a radius of 300 nm, and the Kapitza resistances $\tau_K = 10^{-7}$ m$^2$ K/W.

The first interesting point is that in both Fig. 6 and 7, in the temperature plot, we can identify the jump at the solid-fluid interface corresponding to the Kapitza resistance. Moreover, the (quite constant) temperature within the solid is decreasing with an increasing frequency of the supplied energy. This be-behavior is explained by the last term in Eq. (43), describing the dynamic effect of thermal capacity. Indeed, at high frequencies, the particle's energy absorption cannot follow the rapid temporal variations of the source.

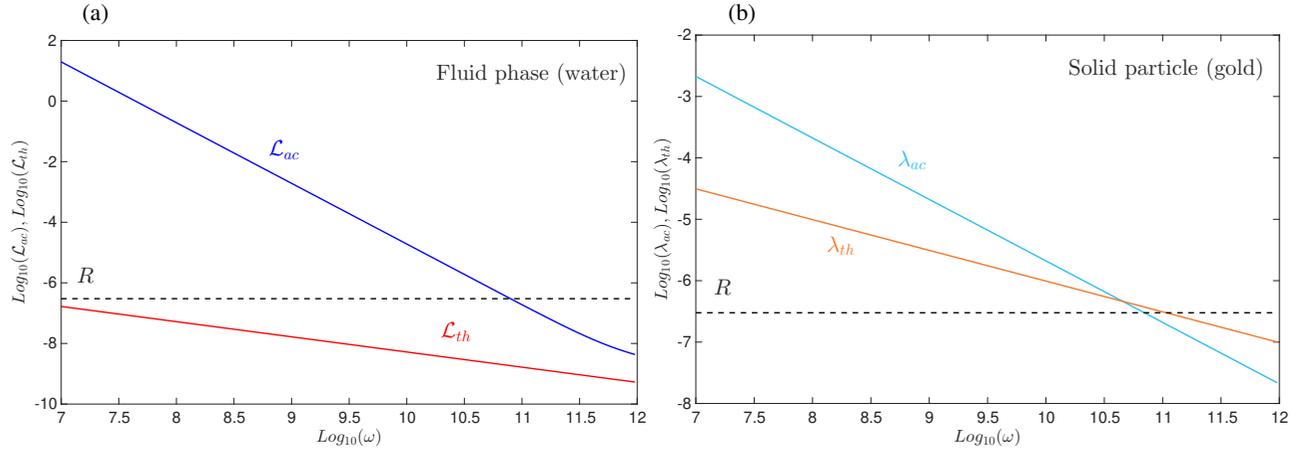

FIG. 5. Panel (a): acoustic $\mathscr{L}_{ac} = -1/\mathfrak{Re}[\vartheta_{0A}]$ and thermal $\mathscr{L}_{th} = -1/\mathfrak{Re}[\vartheta_{0T}]$ penetration lengths in the fluid phase. Panel (b): acoustic $\lambda_{ac} = -2\pi/\mathfrak{Im}[\vartheta_{1A}]$ and thermal $\lambda_{th} = -2\pi/\mathfrak{Im}[\vartheta_{1T}]$ wavelengths within the solid particle. The dashed line corresponding to the radius $R = 300$ nm has been added to the plots for the sake of comparison.

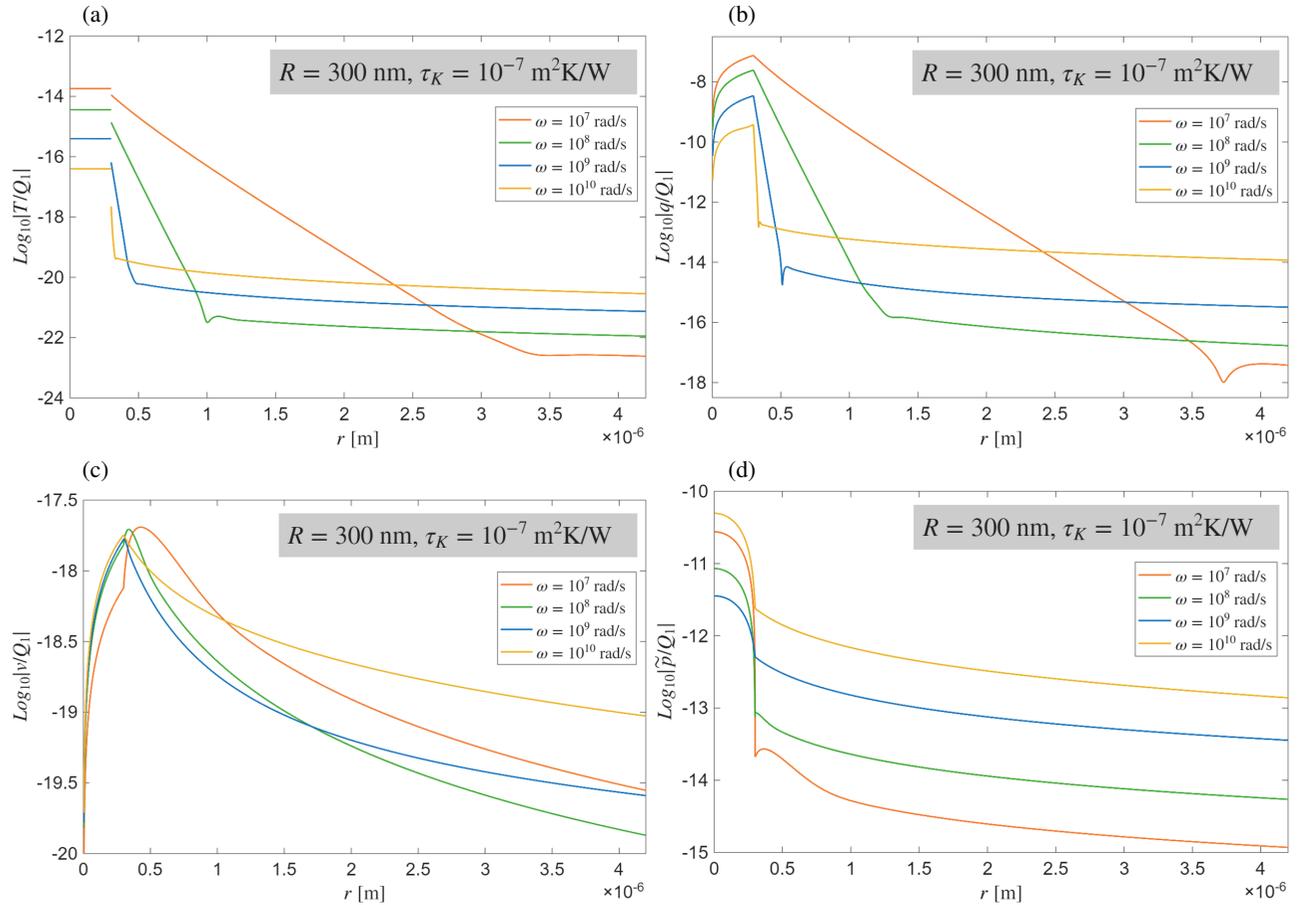

FIG. 6. Spatial distribution of the thermoacoustic fields in the sub-resonance frequency region ($\omega = 10^7, 10^8, 10^8, 10^{10}$ rad/sec). Temperature $T$ (a), heat flux $q$ (b), velocity $v$ (c), and total normal traction $\tilde{p}$ (d) are represented as a function of the radial position $r$. Fields are normalized with respect to the supplied thermal power density $Q_1$, and given in logarithmic scale. We adopted the radius $R = 300$ nm and the Kapitza resistances $\tau_K = 10^{-7}$ m$^2$ K/W.

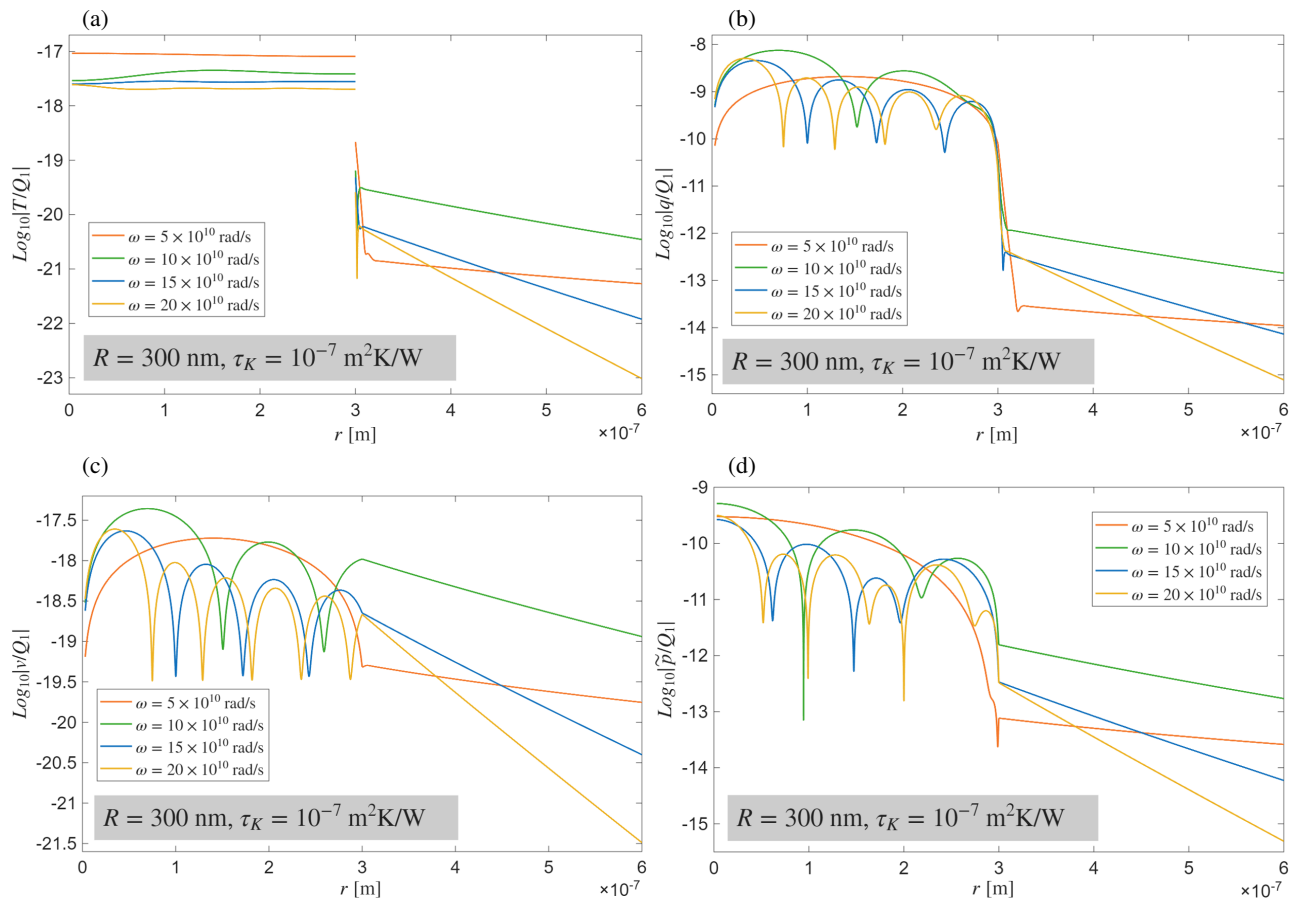

FIG. 7. Spatial distribution of the thermoacoustic fields in the resonance frequency region ($\omega = 5 \times 10^{10}, 10 \times 10^{10}, 15 \times 10^{10}, 20 \times 10^{10}$ rad/sec). Temperature $T$ (a), heat flux $q$ (b), velocity $v$ (c), and total normal traction $\tilde{p}$ (d) are represented as a function of the radial position $r$. Fields are normalized with respect to the supplied thermal power density $Q_1$, given in logarithmic scale. We adopted the radius $R = 300$ nm, and the Kapitza resistances $\tau_K = 10^{-7}$ m$^2$ K/W.

In Figs. 6 and 7, the temperature and heat flux plots, outside the particle, show that the thermal effects penetrate the fluid with a characteristic length $\mathscr{L}_{th}$, inversely proportional to the square root of $\omega$. This result can also be compared with the behavior of $\mathscr{L}_{th}$ versus $\omega$, represented in panel (a) of Fig. 5. Once again, this variation is at the origin of the transition between thermophone and mechanophone mechanisms.

In Fig. 6, we can also observe the behavior of the mechanical fields outside the particle. It is important to remark the the acoustic propagation length $\mathscr{L}_{ac}$ is much larger than the thermal propagation length $\mathscr{L}_{th}$ (at least in the low frequency regime, see Fig. 5), and therefore in the velocity and pressure plots we see long range acoustic fields. Of course, if the distance from the spherical particle is very large, the dissipative phenomena (heat conduction and viscosity) lead to a decreasing trend of these mechanical fields.

The velocity curves in Fig. 6 describe an intricate non-monotonic behavior (in terms of frequency), which can be explained as follows. A key point is the interplay between the thermal penetration depth $\mathscr{L}_{th} = \sqrt{2\kappa_0/(\omega \rho_0 C_{p0})}$, which characterizes the region where thermoacoustic conversion effectively takes place, and the acoustic wavelength in the fluid $\lambda_0 = \frac{2\pi C_0}{\omega}$, where $C_0 = \sqrt{(B_0/\rho_0)(C_{p0}/C_{v0})}$ is the adiabatic sound speed in the same phase. On the one hand, thermoacoustic generation becomes less efficient as the frequency increases, since the active thermal region shrinks according to $\mathscr{L}_{th} \sim 1/\sqrt{\omega}$. This reduction limits the effective volume contributing to pressure generation. On the other hand, the ratio between the thermal depth and the acoustic wavelength scales as $\mathscr{L}_{th}/\lambda_0 \sim \sqrt{\omega}$, so that the thermal source occupies an increasingly significant fraction of the acoustic wavelength as frequency grows. This effect enhances the dynamical coupling between thermal expansion and acoustic radiation. The competition between these two trends naturally leads to the non-monotonic behavior observed in the particle velocity in the near field. At larger radial distances (not shown in the figure), the behavior becomes asymptotically simpler. In the absence of strong dissipative effects, one recovers the scaling $v \sim \omega$, consistent with the harmonic relation $v = i\omega u$, where $u$ is the displacement amplitude. However, at sufficiently high

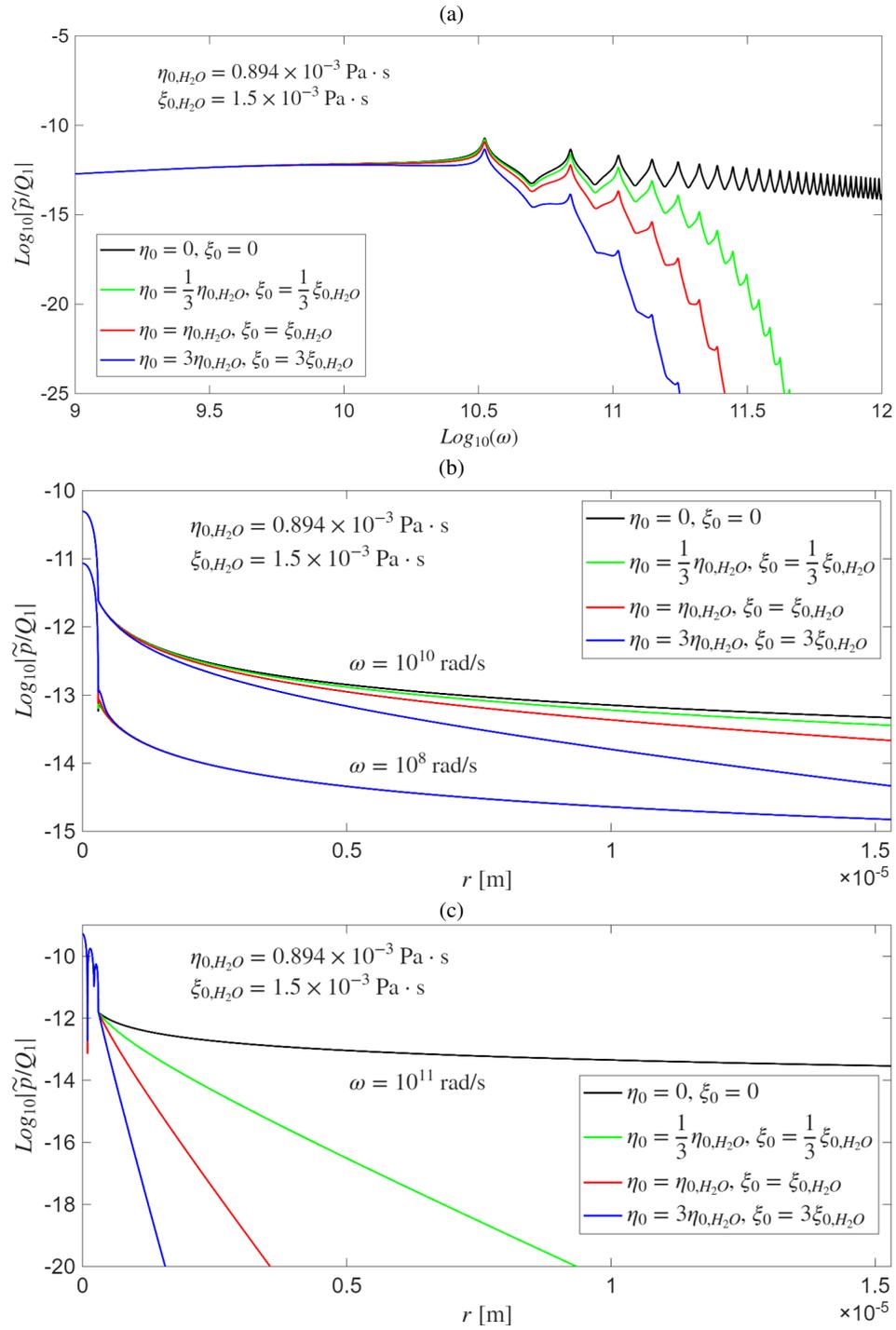

FIG. 8. Effect of the viscosity on the thermoacoustic generation. Panel (a): pressure-frequency response for the system with different values of the viscosity (a case without viscosity, a case with the viscosity of water, and two cases with the viscosity of water divided and multiplied by 3). The fields are measured at the distance $R + 5\mathscr{L}_{th}$, where $\mathscr{L}_{th} = \sqrt{2\kappa_0/(\omega\rho_0 C_{p0})}$ is the thermal length calculated for $\omega = 10$ MHz. Panel (b): spatial distribution (over the range $0 < r < R + 50R$) of the normalized pressure for different values of the viscosity and two values of frequency $\omega = 10^8$ rad/s and $\omega = 10^{10}$ rad/s (sub-resonance region). Panel (c): spatial distribution (over the range $0 < r < R + 50R$) of the normalized pressure for different values of the viscosity and one value of frequency $\omega = 10^{11}$ rad/s (resonance region). In all plots, we adopted the radius $R = 300$ nm, and the Kapitza resistances $\tau_K = 10^{-7}$ m$^2$ K/W.

frequencies this asymptotic trend is progressively modified by viscous and thermal dissipation, which damp the fluid mo- tion and reduce the velocity amplitude. The combined action of geometric features, frequency-dependent thermal penetra-

Acoustic Response of Laser-Excited Nanoparticles 17

tion, and dissipative mechanisms thus accounts for the apparent complexity of the velocity curves.

In Fig. 7, we show the behavior of the thermoacoustic fields in the resonance region. In this case, both the thermal and acoustic propagation lengths are very small, and therefore all the thermoacoustic fields decay exponentially near the solid-fluid interface. These decreasing trends are governed by the length $\mathscr{L}_{th}$ for the temperature and heat flux fields, and by $\mathscr{L}_{ac}$ for the velocity and normal traction fields.

In these plots, the behavior at the interface appears rather intricate, as the solution must simultaneously satisfy all boundary conditions while remaining consistent with the internal resonant structure of the system. This necessarily leads to a nontrivial frequency dependence of the fields in the vicinity of the interface. Despite this apparent complexity, the boundary conditions are fully satisfied: all fields remain continuous across the interface, with the sole exception of the temperature, which exhibits the expected discontinuity induced by the finite interfacial thermal resistance. Moreover, the increasing slope of the curves in the fluid region at higher frequencies clearly reflects the progressive spatial attenuation of the fields. In particular, as the frequency increases, the exponential spatial decay becomes more pronounced, leading to stronger localization of all quantities near the interface.

In this high frequency range (resonance regime), the thermoacoustic fields shown in Fig. 7 exhibit large oscillations (pronounced spatial gradients) within the solid particle since the wavelength is of the same order of magnitude as the particle radius. For the chosen frequencies, it is possible to identify an increasing number of oscillations (from 1 to 4) within the particle. This behavior can be better understood by observing the plot of the acoustic and thermal wavelengths in the solid gold particle, reported in Fig.5 (panel b). It can be seen here that these wavelengths are effectively comparable to or smaller than the radius of the particle in the resonance region.

To conclude the description of the system's behaviour, Fig. 8 shows the effect of viscosity on thermoacoustic generation, bearing in mind that viscosity is often neglected in this type of analysis. Panel (a) shows the frequency response for four different viscosity values. We consider a fluid with no viscosity (black curve), a fluid with the viscosity of water (red curve), and two cases in which the viscosity is one-third (green curve) and three times (blue curve) that of water. Please note that the pressure in these plots, as before, is measured at the distance $R + 5\mathscr{L}_{th}$, where $\mathscr{L}_{th}$ is the thermal length ($\omega = 10$ MHz is fixed to determine $\mathscr{L}_{th}$). It is clear that at high frequencies, acoustic generation is degraded more significantly as viscosity increases (once the measurement position has been set). This behaviour is explained by Eq. (69), which describes the variation of the acoustic penetration $\mathscr{L}_{ac}$ as a function of frequency and the two sources of dissipation, namely viscosity and thermal conduction. In this expression, it is easy to see that $\mathscr{L}_{ac}$ decreases with both frequency and viscosity, consistent with what can be observed in Fig. 8 (panel a).

In panels (b) and (c) of Fig. 8, we show the spatial variation of pressure with different viscosities and frequencies. In particular, panel (b) considers the sub-resonance frequencies $\omega = 10^8$ rad/s and $\omega = 10^{10}$ rad/s, and panel (c) the frequency $\omega = 10^{11}$ rad/s in the resonance region.

In panel (b), for the curves corresponding to $\omega = 10^{10}$ rad/s, it can be seen that the pressure decreases more rapidly as the measuring distance increases when the viscosity is higher. We also observe that for the lower frequency $\omega = 10^8$ rad/s, this phenomenon cannot be observed because the frequency is not high enough, and therefore the curves with different viscosity overlap in the distance range considered (deviations can only be observed at larger distances). In panel (c), we see that at even higher frequencies, the decrease in pressure is even stronger and amplified by higher viscosity values. Furthermore, at these frequencies, pressure oscillations are also observed inside the sphere, typical of the resonance region.

In conclusion, we can say that fluid viscosity plays an essential role in determining the penetration length of the acoustic wave generated by the photothermal effect. If we consider the viscosity of water, this penetration length can vary from a few metres at low frequencies to just a few nanometres at high frequencies. As we have just seen, these values can vary with the viscosity of the medium, and this observation is therefore particularly important for therapeutic or diagnostic applications where a localised effect of the waves used is desired.

## VII. CONCLUSIONS

In this work, we have developed a theoretical framework for the thermoacoustic emission generated by a single laser-heated solid particle embedded in a viscous fluid environment, with reference to the wide range of photoacoustic applications. By solving the fully coupled thermoelastic and thermoacoustic field equations in both phases, and by incorporating the effect of interfacial thermal resistance, we obtained an analytical description of the temperature, heat flux, velocity, and pressure distributions that emerge under harmonic excitation. This formulation provides a unified perspective on the interplay between heat diffusion, thermal expansion, and elastic stress generation at mesoscopic and nanoscopic scales. No approximations were introduced at the level of the continuum model. In particular, the viscosity of both media was fully accounted for, and the effects of fluid viscosity in particular have been studied in great detail to show how this viscosity strongly modifies the acoustic penetration length in the fluid itself. Moreover, the formulation incorporates not only the influence of thermal fields on mechanical behavior, but also the reciprocal effect of mechanical deformations on heat transport—an aspect that is often neglected in standard theoretical treatments. The model is easy to implement numerically and can therefore be readily exploited for biomedical and other applications. We would like to point out that the work is based on the study of frequency response in a stationary harmonic regime. Nevertheless, the results obtained are significant in photoacoustics for studying the temporal response to a given laser pulse by adopting classical techniques based on Fourier series or Fourier transforms. When viewed in the frequency domain, pulsed laser excitation provides access to a broad spectral content, enabling the generation of hypersonic acoustic modes only when ultrashort pulses are employed. In this regime, which far exceeds that of nanosecond excitation, viscous dissipation in the fluid becomes a dominant factor shaping the acoustic response, highlighting the need to explicitly account for dissipative effects in high-frequency thermoacoustic modeling.

Our analysis demonstrates that two distinct physical mechanisms—the thermophone and the mechanophone—govern acoustic emission depending on the frequency regime. At low modulation frequencies, thermal diffusion in the fluid dominates, giving rise to pressure oscillations driven primarily by the periodic heating and cooling of the surrounding medium. At high frequencies, however, thermal penetration depths become small and thermoelastic deformations within the solid become the leading source of acoustic radiation. By selectively suppressing either the solid or the fluid thermal expansion coefficient, we isolated the individual contributions of these two mechanisms, thereby clarifying their roles and identifying the frequency range in which each one prevails.

The general solution presented here also highlights the crucial influence of the Kapitza interfacial resistance. By controlling the rate of heat exchange between the particle and the fluid, interfacial resistance shifts the frequency crossover between thermophone- and mechanophone-dominated regimes, with potentially important consequences for the design of nanoscale photoacoustic sources. Because the Kapitza resistance can be modulated through material choice, surface chemistry, or functionalization, it represents an accessible and powerful tuning parameter for optimizing device performance.

Beyond mapping the thermophone–mechanophone transition, the analytical framework developed in this study allows for the prediction of resonance frequencies in the high-frequency regime, where the particle's intrinsic elastic modes significantly enhance the generated acoustic field. These resonances, and their sensitivity to particle radius and material parameters, offer further opportunities for tailoring acoustic response for specific applications. Another crucial aspect is the interaction between the applied frequency (or pulse duration) and the fluid viscosity, which determines the length of acoustic penetration into the fluid and therefore implicitly controls the therapeutic and diagnostic functions of the system under examination.

In the present work, we have restricted the analysis to a single spherical particle in order to isolate the fundamental thermo–acoustic mechanisms and to derive an exact analytical solution of the coupled problem. This approach allows us to clearly disentangle the interplay between thermal diffusion, elastic response, and acoustic radiation, without additional collective effects. Nevertheless, the proposed framework can be extended to a population of particles in a natural way (see Fig. 1, left panel). In the dilute regime, where the interparticle distance is large compared to both the thermal diffusion length and the acoustic wavelength, thermal overlap and multiple-scattering effects remain negligible. Under these conditions, owing to the linearity of the governing equations, the overall response of the suspension can be approximated as the linear superposition of the fields generated by the individual particles. Accordingly, within the limits of weak interparticle coupling (low volume fraction and negligible multiple scattering), the single-particle solution derived here constitutes the fundamental building block for describing dispersed systems. Only at higher concentrations, where collective thermal or acoustic interactions become significant, would a more refined effective-medium or multiple-scattering description be required.

Overall, the present work provides a rigorous foundation for understanding photo-thermo-acoustic generation from laser-activated particles, bridging thermal, mechanical, and acoustic perspectives into a single coherent model. The resulting insights can inform the rational design of broadband nanotransducers and enable more efficient exploitation of photo-thermo-acoustic phenomena in emerging biomedical and technological applications, including localized imaging, targeted hyperthermia, and multifunctional theranostic systems.

## ACKNOWLEDGMENTS

Stefano Giordano acknowledges support from a Visiting Professor Grant of the University of Cagliari. Michele Brun acknowledges support from a Visiting Professor Grant of the University of Lille. The authors sincerely thank the reviewer for the constructive comments, which helped improve the clarity and quality of the manuscript.

## AUTHOR DECLARATIONS

**Conflict of Interest**

The authors have no conflicts to disclose.

**Author Contributions**

Stefano Giordano: Conceptualization (equal); Formal analysis (equal); Investigation (equal); Methodology (equal); Writing – review & editing (equal). Michele Diego: Conceptualization (equal); Formal analysis (equal); Investigation (equal); Methodology (equal); Writing – review & editing (equal). Francesco Banfi: Conceptualization (equal); Formal analysis (equal); Investigation (equal); Methodology (equal); Writing – review & editing (equal). Michele Brun: Conceptualization (equal); Formal analysis (equal); Investigation (equal); Methodology (equal); Writing – review & editing (equal).

## DATA AVAILABILITY

No data was used for the research described in the article.

## Appendix A: Differential operators under spherical symmetry

If a scalar function (e.g., the pressure $p$) is characterized by a spherical symmetry $p = p(r)$, the gradient operator simplifies as follows

$$\vec{\nabla} p = \frac{\partial p}{\partial x_i} \vec{e}_i = \frac{dp}{dr} \frac{\partial r}{\partial x_i} \vec{e}_i = \frac{x_i}{r} \frac{dp}{dr} \vec{e}_i = \frac{\vec{r}}{r} \frac{dp}{dr}, \quad (A1)$$

where $\vec{e}_1$, $\vec{e}_2$, and $\vec{e}_3$ are the unit vectors along the axes of the orthogonal frame, $\vec{r} = (x_1, x_2, x_3) = x_i \vec{e}_i$, and $r = \sqrt{x_1^2 + x_2^2 + x_3^2}$. We can also determine the Laplacian operator as

$$\nabla^2 p = \frac{\partial}{\partial x_i} \frac{\partial}{\partial x_i} p = \frac{\partial}{\partial x_i} \left( \frac{x_i}{r} \frac{dp}{dr} \right) = \frac{1}{r^2} \frac{d}{dr} \left( r^2 \frac{dp}{dr} \right). \quad (A2)$$

If a vector field (e.g., the velocity field) exhibits spherical symmetry, it can be written as

$$\vec{v} = \frac{\vec{r}}{r} v(r), \quad v_i = \frac{x_i}{r} v(r). \quad (A3)$$

The following derivative (without summation over $i$)

$$\frac{\partial v_i}{\partial x_i} = \frac{v}{r} + \frac{x_i^2}{r^2} \frac{dv}{dr} - v \frac{x_i^2}{r^3} \quad (A4)$$

is useful to obtain the simplified divergence operator

$$\vec{\nabla} \cdot \vec{v} = 2 \frac{v}{r} + \frac{dv}{dr} = \frac{1}{r^2} \frac{d}{dr} \left( r^2 v \right). \quad (A5)$$

We can also define the Laplacian of $\vec{v}$ as

$$\nabla^2 \vec{v} = \nabla^2 \left( \frac{\vec{r}}{r} v \right), \quad (A6)$$

or in components

$$\left( \nabla^2 \vec{v} \right)_j = \frac{\partial}{\partial x_i} \left[ \frac{\partial}{\partial x_i} \left( \frac{x_j}{r} v \right) \right], \quad (A7)$$

with summation over $i$. By performing the derivatives, we get

$$\left( \nabla^2 \vec{v} \right)_j = \frac{\partial}{\partial x_i} \left[ \delta_{ij} \frac{v}{r} + \frac{x_i x_j}{r^2} \frac{dv}{dr} - x_i x_j \frac{v}{r^3} \right]$$
$$= \frac{x_j}{r} \left( \frac{d^2 v}{dr^2} + \frac{2}{r} \frac{dv}{dr} - 2 \frac{v}{r^2} \right), \quad (A8)$$

and therefore

$$\nabla^2 \vec{v} = \frac{\vec{r}}{r} \left( \frac{d^2 v}{dr^2} + \frac{2}{r} \frac{dv}{dr} - 2 \frac{v}{r^2} \right)$$
$$= \frac{\vec{r}}{r} \frac{d}{dr} \left[ \frac{1}{r^2} \frac{d}{dr} \left( r^2 v \right) \right]. \quad (A9)$$

Similarly, we can also calculate the gradient of the divergence as

$$\vec{\nabla} \left( \vec{\nabla} \cdot \vec{v} \right) = \vec{\nabla} \left( 2 \frac{v}{r} + \frac{dv}{dr} \right), \quad (A10)$$

from which an arbitrary component is given by

$$\left[ \vec{\nabla} \left( \vec{\nabla} \cdot \vec{v} \right) \right]_i = \frac{\partial}{\partial x_i} \left( 2 \frac{v}{r} + \frac{dv}{dr} \right)$$
$$= \frac{x_i}{r} \left( \frac{d^2 v}{dr^2} + \frac{2}{r} \frac{dv}{dr} - 2 \frac{v}{r^2} \right), \quad (A11)$$

and therefore with spherical symmetry

$$\vec{\nabla} \left( \vec{\nabla} \cdot \vec{v} \right) = \nabla^2 \vec{v} = \frac{\vec{r}}{r} \frac{d}{dr} \left[ \frac{1}{r^2} \frac{d}{dr} \left( r^2 v \right) \right]. \quad (A12)$$